\documentclass[preprint]{aastex}
\usepackage{graphicx}
\usepackage{natbib}
\begin{document}

\title{Aegaeon (Saturn LIII), a G-ring object
\footnote{This paper is dedicated to the memory of Kevin Beurle.} }
\author{M.M. Hedman$^a$, N.J. Cooper$^b$, C.D.Murray$^b$, 
K. Beurle$^b$, M.W. Evans$^{b,a}$,\\ M.S. Tiscareno$^a$, J. A. 
Burns$^{a,c}$}
\affil{
	$^a$ Department of Astronomy, Cornell University, Ithaca NY 14853 USA \\
	$^b$ Queen Mary University of London, 
		Astronomy Unit, School of Mathematical Sciences, Mile End Road, London E1 4NS, UK \\
	$^c$ Department of Theoretical and Applied Mechanics, 
	Cornell University, Ithaca NY 14853 USA \\}

\maketitle

\bigskip

\bigskip

\bigskip

Proposed Running Head: Aegaeon (Saturn LIII), a G-ring object

Keywords: Saturn, Satellites; Saturn, rings; Satellites, dynamics; Resonances, orbital

\pagebreak

{\bf Abstract:} Aegaeon (Saturn LIII, S/2008 S1) is a small satellite of Saturn 
that orbits within a bright arc of material near the inner edge of 
Saturn's G ring. This object was observed
in 21 images with Cassini's Narrow-Angle Camera between 
June 15 (DOY 166), 2007 and February 20 (DOY 51), 2009. If Aegaeon
has similar surface scattering properties as other 
nearby small Saturnian satellites (Pallene, Methone and Anthe), 
then its diameter is approximately 500 m. 
Orbit models based on numerical integrations of the full equations of 
motion show that Aegaeon's orbital motion is strongly influenced
by multiple resonances with Mimas. In particular, like the G-ring arc it inhabits, 
Aegaeon is trapped in the 7:6 corotation eccentricity resonance with Mimas. 
Aegaeon, Anthe and Methone therefore form a distinctive
class of objects in the Saturn system: small moons
 in co-rotation eccentricity resonances with Mimas
associated with arcs of debris. Comparisons among these
different ring-arc systems reveal that Aegaeon's orbit is closer to
the exact resonance than Anthe's and Methone's orbits are.
This could indicate that Aegaeon has undergone significant orbital evolution via
its interactions with the other objects in its arc, which would
be consistent with the evidence that Aegaeon's mass is much 
smaller relative to the total mass in its arc than Anthe's and Methone's masses are.

\section{Introduction}

Beginning in early 2004, images from the cameras onboard 
the Cassini spacecraft revealed the existence of several 
previously unknown small Saturnian satellites: Methone, Pallene,
Polydeuces, Daphnis and Anthe \citep{Porco05, Murray05, Spitale06, Cooper08}. 
Two of these moons --Anthe and Methone-- are in mean-motion 
resonances with Saturn's moon Mimas. Specifically, they 
occupy the 10:11 and 14:15 co-rotation eccentricity resonances, 
respectively \citep{Spitale06, Cooper08, Hedman09}. 
Both of these moons are also embedded in very faint, 
longitudinally-confined  ring arcs \citep{Roussos08, Hedman09}. 
This material probably represents 
debris that was knocked off the relevant moons at low velocities and thus
remains trapped in the same co-rotation resonance as its source body.

Images from Cassini 
also demonstrated that a similar arc of material exists within 
Saturn's G ring, around 167,500 km from Saturn's center \citep{Hedman07}. 
Images of this structure taken over the course of several years showed that it was 
also confined by a (7:6) corotation eccentricity resonance with Mimas. 
Furthermore, in-situ measurements of the plasma environment in the vicinity 
of the arc suggested that it contains a significant amount of mass in particles 
larger than the dust-sized grains that are the dominant source of 
scattered light observed in images \citep{Hedman07}.

\nocite{Porco09}
In late 2008, during Cassini's Equinox Mission (2008-2010), 
images of the arc taken at lower phase angles and 
higher resolutions than previously possible revealed a small, discrete
object.  Since the object was most visible at low
phase angles and could be tracked over a period of roughly 600 days, 
it is almost certainly not a transient clump of dust but instead a tiny moonlet that represents
the largest of the source bodies populating the arc.  The discovery of this
object was therefore announced in an IAU circular, where it was  
designated  S2008/S1 (Porco {\it et al.} 2009). More recently
the International Astronomical Union has given it 
the name Saturn LIII/Aegaeon. As will be shown below, Aegaeon, like Anthe and 
Methone, occupies a corotation eccentricity resonance
with Mimas, and all three of these small moonlets are associated with 
arcs of debris. These three objects
therefore represent a distinct class of satellites and comparisons among the
ring-moon systems have the potential to illuminate the connection
between moons and rings.

Section 2 below describes the currently available images of Aegaeon and how 
they are processed to obtain estimates of the brightness and position of
this object. Section 3 presents a preliminary analysis of the photometric data, 
which indicate that this object is approximately 500 m in diameter. Section 4 
describes the
orbital solutions to the astrometric data, which demonstrate that Aegaeon's
orbit is indeed perturbed by the 7:6 corotation eccentricity resonance with Mimas.
However,we also find that a number of other resonances, including the
7:6 Inner Lindblad Resonance, strongly influence Aegaeon's orbital motion.
Finally, Section 5 compares the various resonantly-confined
moon/ring-arc systems to one another in order to clarify the relationship 
between Aegaeon and the G ring.

\section{Observational Data}

The images discussed here were obtained with the 
Narrow-Angle Camera (NAC) of the Imaging Science
Subsystem (ISS) onboard the Cassini spacecraft \citep{porco04}.
All images were initially processed using the CISSCAL 
calibration routines \citep{porco04} that remove backgrounds,
flat-field the images, and convert the raw data numbers 
into $I/F$, a standardized measure of reflectance. $I$ is
the intensity of the scattered radiation while $\pi F$ is the
solar flux at Saturn, so $I/F$  is a unitless quantity that
equals unity for a perfect Lambert surface viewed at normal 
incidence.

\subsection{Image Selection}

\begin{figure}
\centerline{\resizebox{5in}{!}{\includegraphics{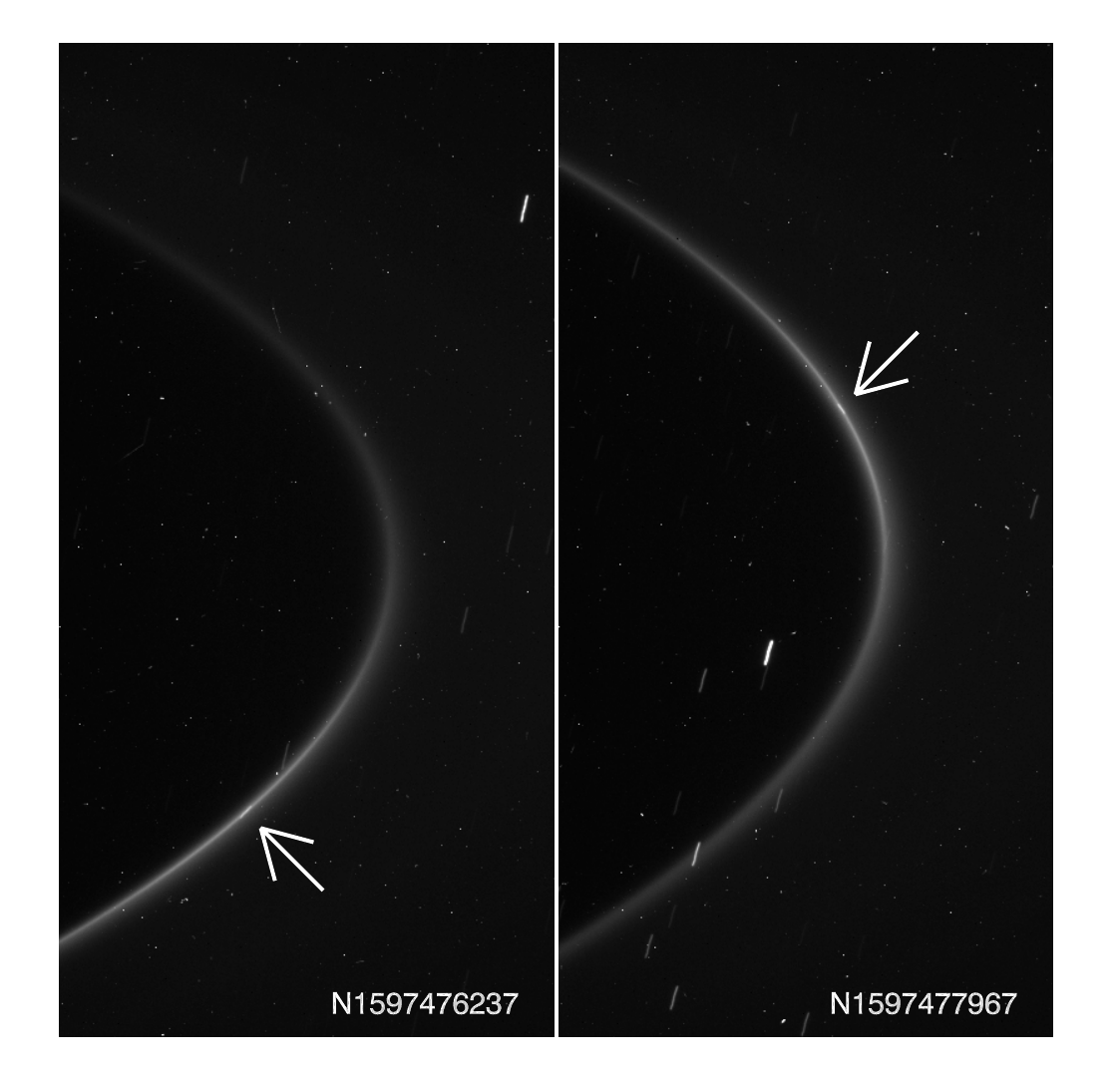}}}
\caption{The pair of images taken on August 15 (DOY 228), 2008 in which 
Aegaeon was first noticed. The arrows point to this object, which appears as
a small streak within the core of the G ring due to its orbital motion through
the field of view over the course of these long-exposure images.
Both images are rotated so that  Saturn's north pole would point towards the
top of the page.}
\label{discim}
\end{figure}

The object was first noticed in two images taken on August 15
(Day-of-year 228), 2008
(see Fig.~\ref{discim}).
These images were part of a sequence designed to image
the arc in the G ring for the purposes of refining its orbit.
Compared with previous imaging of the G-ring arc, the images
used in this campaign were taken at lower phase angles and
had better spatial resolution. This was more a result of the constraints
imposed by the orbit geometry than a conscious effort to search for discrete 
objects in this region. 
When these images were taken, Cassini was in a highly inclined 
orbit with the ascending node near apoapse on 
the sunward side of the planet close to Titan's orbit.
During these ring-plane crossings, the faint rings could be
imaged at high signal-to-noise, and the low-phase angles
were considered desirable because this geometry
was comparatively rarely observed prior to this time.
However,  this geometry  also turned out to be useful for
detecting small objects in the G ring. 

Two images from this sequence (Fig.~\ref{discim}) contained the core of the arc
and also showed a short, narrow streak  in the G ring. 
The streaks are aligned with the local orbital motion of the arc and 
are clearly not aligned with the streaks associated with stars in the 
field of view. The lengths of the streaks are consistent with the expected movement
of an object embedded in the arc over the exposure time, and the positions of the streaks in the two images are consistent with such an object's motion over the $\sim$30 minutes
between the two images. 

\begin{figure}
\centerline{\resizebox{5.0in}{!}{\includegraphics{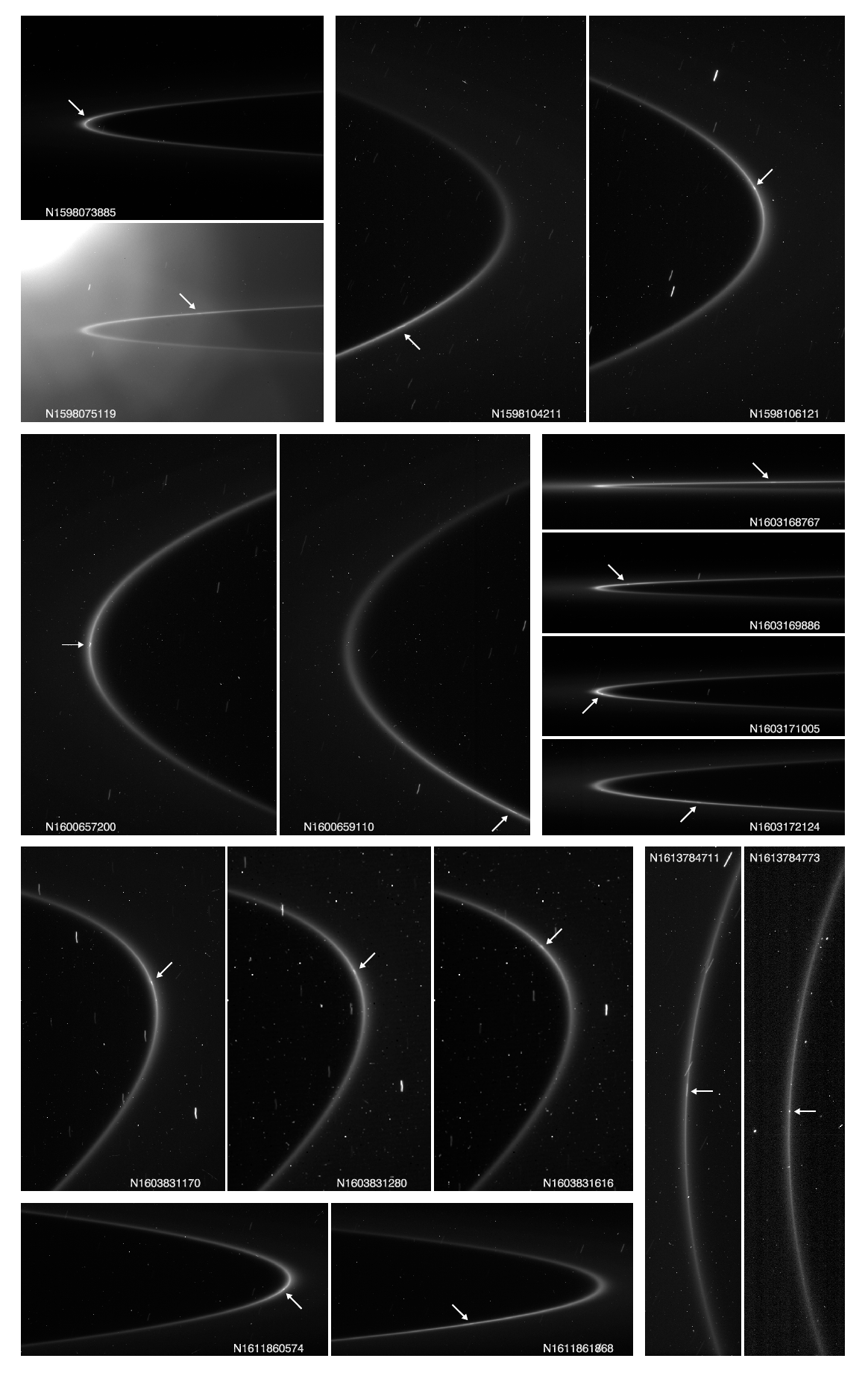}}}
\caption{Other low-phase, high-resolution images of Aegaeon
obtained from late 2008 through February 20 (DOY 051), 2009. In each
image the object's location is highlighted with an arrow. All 
images are rotated so Saturn's north pole would point upwards.
Note the bright feature in the upper left corner of image N1598075119
is due to Tethys being in the camera's  field of view.}
\label{postim}
\end{figure}

\begin{figure}[htbp]
\centerline{\resizebox{5.0in}{!}{\includegraphics{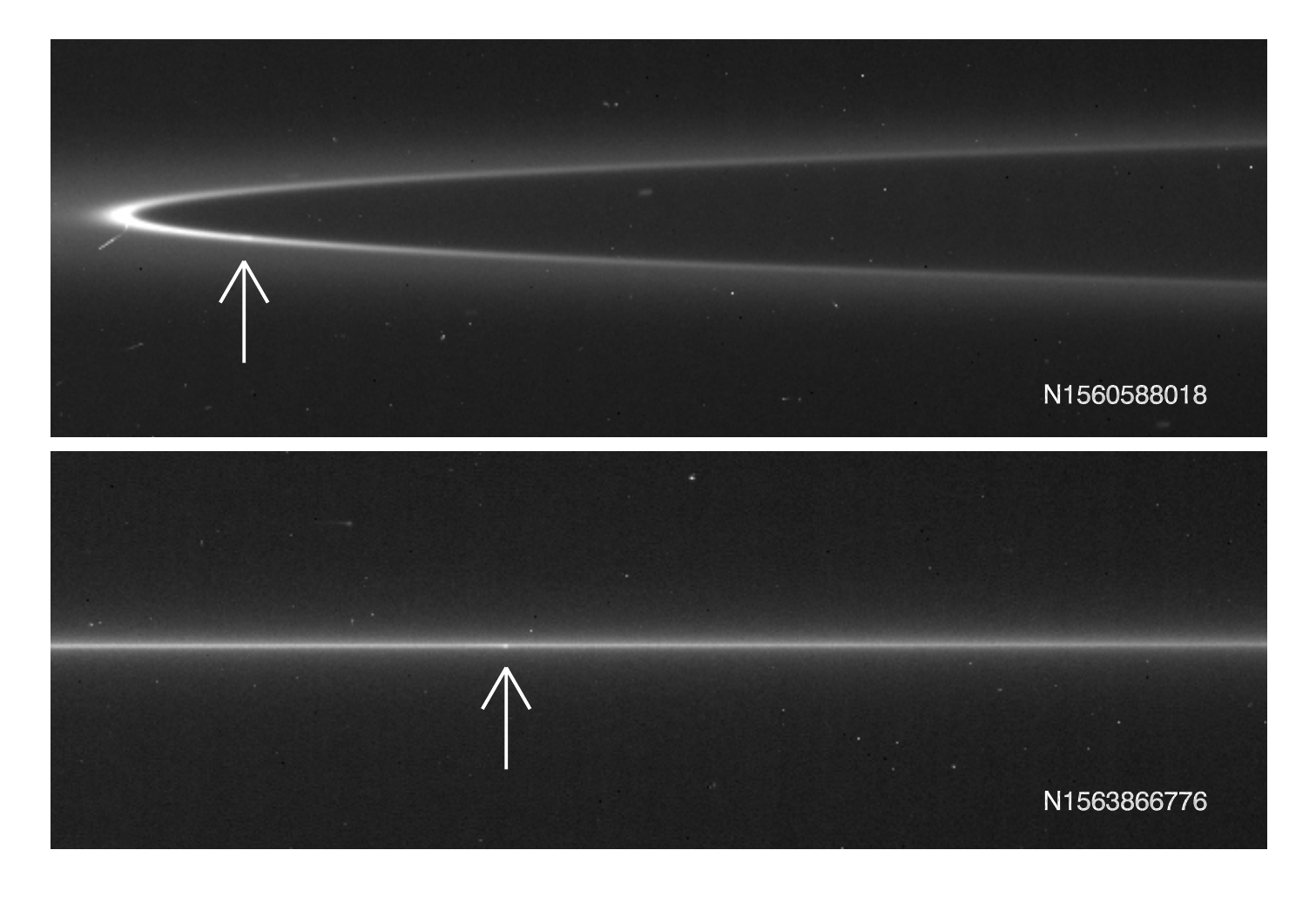}}}
\caption{The only two clear images of Aegaeon
obtained prior to mid-2008 found to date. In each
image the location of the object is highlighted with an arrow. Both 
images are rotated so Saturn's north pole would point roughly up.}
\label{preim}
\end{figure}

Since this sequence was part of a larger campaign designed
to track the arc and refine its orbit, this object was quickly recovered
in subsequent image sequences targeted at the arc with comparable viewing 
geometries, yielding 17 additional images of the object (Fig.~\ref{postim}). 
With these data, a preliminary orbit fit was used to
search for earlier images of the object. However, only two images
from the prime mission turned out to provide clear detections of 
Aegaeon (Fig.~\ref{preim}). This paucity of pre-discovery images is because this
object is both extremely faint and embedded in the G-ring arc. 
While the object's faintness means that it cannot be clearly detected in
images where the exposure times are too
short, its proximity to the G-ring arc means that its signal cannot
be isolated if the image resolution is too low or the phase angle
is too high.

Table~\ref{imgtab} lists the  21 NAC images used in this analysis, which
are all the images prior to February 20 (DOY 051), 2009 in which Aegaeon has been 
securely identified. These images cover a time interval of almost 
600 days and a range of phase angles from 13$^\circ$ to 43$^\circ$.

\subsection{Image Data Reduction}

Since Aegaeon is not resolved in any of the images listed
in Table~\ref{imgtab}, the only data we can extract
from each image are its position in the field of view and its
total integrated brightness. However, estimating even these
parameters from these images is challenging because the light
from Aegaeon is smeared out into a streak and because
the light from the object must be isolated from the background 
signal from the G ring arc. The following procedures were
used to obtain
the required photometric and astrometric data.

In order to isolate the moon's signal from that of 
the G ring, each image was first roughly navigated based on stars
within the field of view. Then, the radius and longitude in the
ringplane observed by each pixel was computed.  Based
on visual inspection of the image, a region of the
image containing the arc was selected (in general these regions
are 10-20 pixels across). A second region extending 10 pixels
beyond this zone on either side along the ring 
was then used to construct a radial profile of the G ring and arc in the
vicinity of the moon. A background based on this profile was then subtracted 
from the pixels in the selected region, which removes the signal from the 
G ring and arc, leaving behind only  the signal from Aegaeon itself. 

Two images were handled slightly differently because they were
taken in a nearly-edge-on viewing geometry (N1563866776 and
N1603168767). In these cases instead of computing radius and 
longitude for each pixel, we compute the radius and vertical 
height above the ringplane and remove a vertical brightness 
profile from the region around the object.

After separating Aegaeon's signal from the G ring,
the total brightness of the object in each image is 
estimated in terms of an effective area, which is the equivalent 
area of material with $I/F=1$ required to account for the 
observed brightness:
\begin{equation}
A_{eff}=\sum_x\sum_y I/F(x,y)*\Omega_{pixel}*R^2,
\label{aeff}
\end{equation}
where $x$ and $y$ are the line and sample numbers of the
pixels in the selected region,$I/F(x,y)$ is the (background-subtracted) 
brightness of the streak in the $x,y$ pixel,
$\Omega_{pixel}=(6\mu$rad$)^2$ is the assumed solid angle 
subtended by a NAC pixel, and $R$ is the distance between the
spacecraft and the object during the observation. The assumed values for
$R$ (given in Table~\ref{imgtab}) are based on the best current orbital solution
(see below).

Similarly, the object's mean position in the field of view was determined
by computing the coordinates (in pixels) of the streak's center of light 
$x_c$ and $y_c$:
\begin{equation}
x_c=\frac{\sum_x\sum_y x*I/F(x,y)}{\sum_x\sum_y I/F(x,y)},
\end{equation}
\begin{equation}
y_c=\frac{\sum_x\sum_y y* I/F(x,y)}{\sum_x\sum_y I/F(x,y)}.
\end{equation}
For purposes of deriving the object's orbit,
these estimates of Aegaeon's position within the camera's
field of view are converted into estimates of its
right ascension and declination on the sky as seen by Cassini.
This is accomplished by comparing the center-of-light coordinates
of Aegaeon to the center-of-light coordinates of various stars 
in the field of view. 

Table~\ref{imgtab} lists all derived parameters
for each of the relevant images.

\section{Photometric analysis and the size of Aegaeon}

Table~\ref{imgtab} includes 19 measurements of 
Aegaeon's brightness through the NAC's 
clear filters over a range of phase angles between
13$^\circ$ and 43$^\circ$.\footnote{Two images (N1603831280 and
N1603831616) were obtained using the RED ($\lambda_{eff}$= 649 nm) 
and IR1 ($\lambda_{eff}$=751 nm) filters, respectively.
While Aegaeon's brightness is the same at both these
wavelengths at the 5\% level, it is premature
to make any definite conclusions about Aegaeon's  color 
based on such limited data.}
In the absence of disk-resolved images of 
this object, these photometric data provide the only 
basis for estimating its size.

For the above range of phase angles $\alpha$, the  effective area $A_{eff}$ of a 
spherical object is usually well approximated by the
following form:
\begin{equation}
A_{eff}=p_{eff}A_{phys}10^{-\beta\alpha/2.5},
\label{phaseeq}
\end{equation}
where $A_{phys}$ is the physical cross-sectional
area of the object, $p_{eff}$ is the effective
geometric albedo (neglecting the opposition surge)
and $\beta$ is the linear phase coefficient \citep{Veverka77}.
Even if the object is not spherical,
we still expect that $<A_{eff}(\alpha)>$ --the
effective area at a given phase angle averaged
over object orientations-- will have the same basic form:
\begin{equation}
<A_{eff}>=p_{eff}<A_{phys}>10^{-\beta\alpha/2.5}
\label{phaseeq2}
\end{equation}
where $<A_{phys}>$ is the average physical 
cross-section of the object.

Fitting the photometric data over a 
sufficiently broad range of phase angles
to Equation~\ref{phaseeq2} can provide
estimates of the linear phase coefficient  $\beta$  and the 
product $p_{eff}<A_{phys}>$. However, to
convert the latter into an estimate of the 
object's size requires additional information
about $p_{eff}$, which can be obtained from comparisons with 
similar objects. For Aegaeon, the
best points of comparison are Pallene, Methone and Anthe,
three small Saturnian moons whose orbits
lie between those of Mimas and Enceladus.
These moons are the closest in size
to Aegaeon and are in similar environments (Pallene, Methone, Anthe and 
Aegaeon are all embedded in faint rings or arcs of material). 
 
To quantitatively compare the photometric
characteristics of these various moons,
we computed the effective areas $A_{eff}$
of Pallene, Methone and Anthe from
a series of images taken over a similar
range of phase angles as the Aegaeon images. 
Tables~\ref{paltab},
~\ref{mettab} and ~\ref{anttab} list the
images of Pallene, Methone, and Anthe
used in this analysis. Since
the goal here is to make comparisons between
different moons and not to do a complete
photometric analysis of these objects,  
the images used in the current study 
are only a selected subset
of NAC clear-filter images that were expected
to give the most reliable brightness data based
on the spacecraft range and exposure duration.
All these images were taken from within about 2 
million kilometers of the target
moon and had exposure durations that were
long enough for the
moon's signal to be measured accurately but short enough
that there was no chance of saturation.
 
For each image, we computed the total integrated brightness 
in  a 14-by-14 pixel wide zone containing the moon
above the average background level in a 5-pixel wide annulus
surrounding the selected region. These total brightness
measurements were then converted into effective
areas using the range between the spacecraft and the 
moon as described in Equation~\ref{aeff}.

\begin{figure}[htbp]
\centerline{\resizebox{5in}{!}{\includegraphics{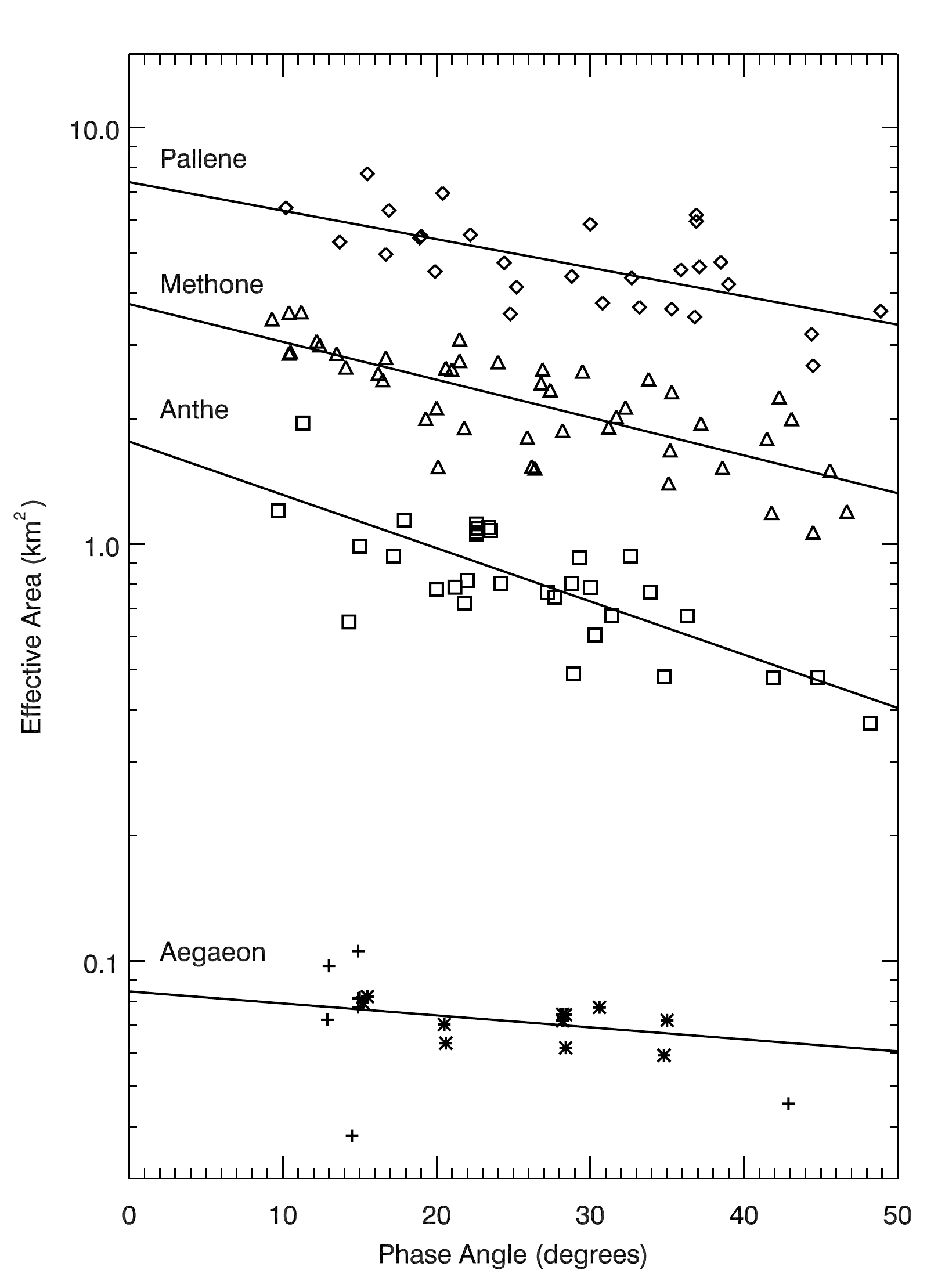}}}
\caption{The effective areas of Pallene, Methone, Anthe and
Aegaeon as  functions of phase angle. Note that  most of 
the scatter in the Pallene, Methone and Anthe around the
trend-line are due to variations in the
orientation of the moon relative to the spacecraft. Two
sub-sets of the Aegaeon data are highlighted. The
stars are data from images with ring opening angles
above 1$^\circ$, which are considered more reliable
than those obtained at lower ring-opening angles
(marked as plusses) where the contrast of the object
against the ring is weaker. Since the scatter in the data 
for each moon is dominated by systematic effects,
statistical error bars are not included in this plot.} 
\label{moonarea}
\end{figure}

Figure~\ref{moonarea} shows the resulting
estimates of $A_{eff}$ as a function of phase angle
for Pallene, Methone, Anthe and Aegaeon.
The data for Pallene, Methone and Anthe all
show significant scatter around the main trends.
In all three cases, this scatter can be attributed to
variations in the orientation of a non-uniform or non-spherical object relative to 
the spacecraft (As will be discussed in a future work, 
all three moons appear to have significant
ellipticities with the long axis pointing towards Saturn). The Aegaeon
data are divided into two groups in this plot based on
whether the observation had a ring opening
angle $|B|$  greater or less than 1$^\circ$. Because the 
contrast of the moon against the background G ring
is reduced at lower ring opening angles due to
the increased surface brightness of the ring material, the
$|B|>1^\circ$ data are considered to be more reliable
measurements of $A_{eff}$. 

Despite the scatter, it is clear that the data from all
four objects can be fit to a mean trend of the
form given in Equation~\ref{phaseeq2}. The lines
in the plots show the resulting best-fit trend, while
Table~\ref{moontab} gives the resulting fit parameters
(note only the $|B|>1^\circ$ data are used for the Aegaeon
fit). Because the scatter
in the data points from each moon is not random error, but
instead  systematic variations associated with viewing geometry, 
error bars on these parameters are not reported here.

The phase coefficients of Anthe, Pallene and Methone
are reasonable values for small airless objects (compare with 
values for asteroids in Bowell and Lumme 1979), while
the coefficient for Aegaeon is somewhat on the low side, 
which may be because a residual unsubtracted  G-ring signal 
adds a slightly forward-scattering component to its phase 
curve. Alternatively, Aegaeon may have a smoother surface
than the other moons \citep{Veverka71}.
\nocite{BL79}

Of all of these moons, only Pallene has been observed
with sufficient resolution to obtain a well-defined mean radius
of 2.2.$\pm$ 0.3 km \citep{Porco07}. Given the observed value
of $p_{eff}<A_{phys}> = 7.38$ km$^2$, this would imply that
$p_{eff} =0.49$ for this moon. Assuming that all four objects have 
roughly the same geometric albedo, we obtain estimates of the
mean radii of Methone and Anthe of 1.6 km and 1.1 km, 
respectively. The estimated size of Methone matches the estimate
derived from crudely resolved images (1.6 $\pm$ 0.6 km, Porco {\it et al.} 2007), 
and the radius of Anthe matches previous estimates based on its brightness relative
to Pallene \citep{Cooper08}. 
Applying this same albedo to Aegaeon suggests a radius of 240 m.
Assuming geometric albedos between 0.1 and 1.0 gives a range
of radii between 160 and 520 m, so although the size of the object is
still uncertain, it is almost certainly less than 1 km across.

\section{Orbital Solutions}

The methodology used to derive the orbital solution for Aegaeon follows the
same basic procedures used by \citet{Cooper08} with Anthe and \citet{Murray05}
with Polydeuces.  As in those works,
the solution is computed in a planetocentric reference frame
where the $x$-axis corresponds to the direction
of the ascending node of Saturn's equatorial plane on the equator of the International
Celestial Reference Frame (ICRF); the $z$ axis is directed along Saturn's
spin axis at epoch (pointing north); and the $y$-axis is orthogonal to $x$ and $z$ 
and oriented as required to produce a right-handed coordinate system.
The chosen epoch for the orbital solution is 2008-228T06:45:07.972 UTC
(the time of the first discovery image). The assumed values for Saturn's pole 
position and gravitational field are given in Table~\ref{satparams}, 
while Table~\ref{kernels} lists the SPICE kernels \citep{Acton96} used
in the orbit determination and numerical modeling.

As with Anthe and other small Saturnian satellites, the orbit of Aegaeon
cannot be accurately fit with a simple precessing elliptical model (see below).
Thus the data were fit to a numerical integration of the full equations of 
motion in three dimensions,  solving for the initial state of Aegaeon at epoch.
This model included perturbations from the Sun, Saturn, Jupiter, 
the eight major satellites of Saturn (Mimas, Enceladus, Tethys, 
Dione, Rhea, Titan, Hyperion and Iapetus), as well as 
the smaller moons Prometheus, Pandora, Janus and Epimetheus.
The forces from these perturbers were calculated using position
vectors extracted from the JPL ephemerides listed in Table~\ref{kernels}.
These position vectors were rotated from the ICRF frame to 
the saturn-centric reference frame using the pole position
given in Table~\ref{satparams}, obtained by precessing the
pole position of \citet{Jacobson04} to the chosen epoch, using rates of
-0.04229$^\circ$/cy in right acension and  -0.00444$^\circ$/cy in declination 
\citep{Jacobson04}. Terms up to $J_6$ in Saturn's gravitational field were taken into
account. The adopted $GM$ values for the satellites, etc. are given
in Table~\ref{pertparams}.

\begin{figure}[tbp]
\centerline{\resizebox{6in}{!}{\includegraphics{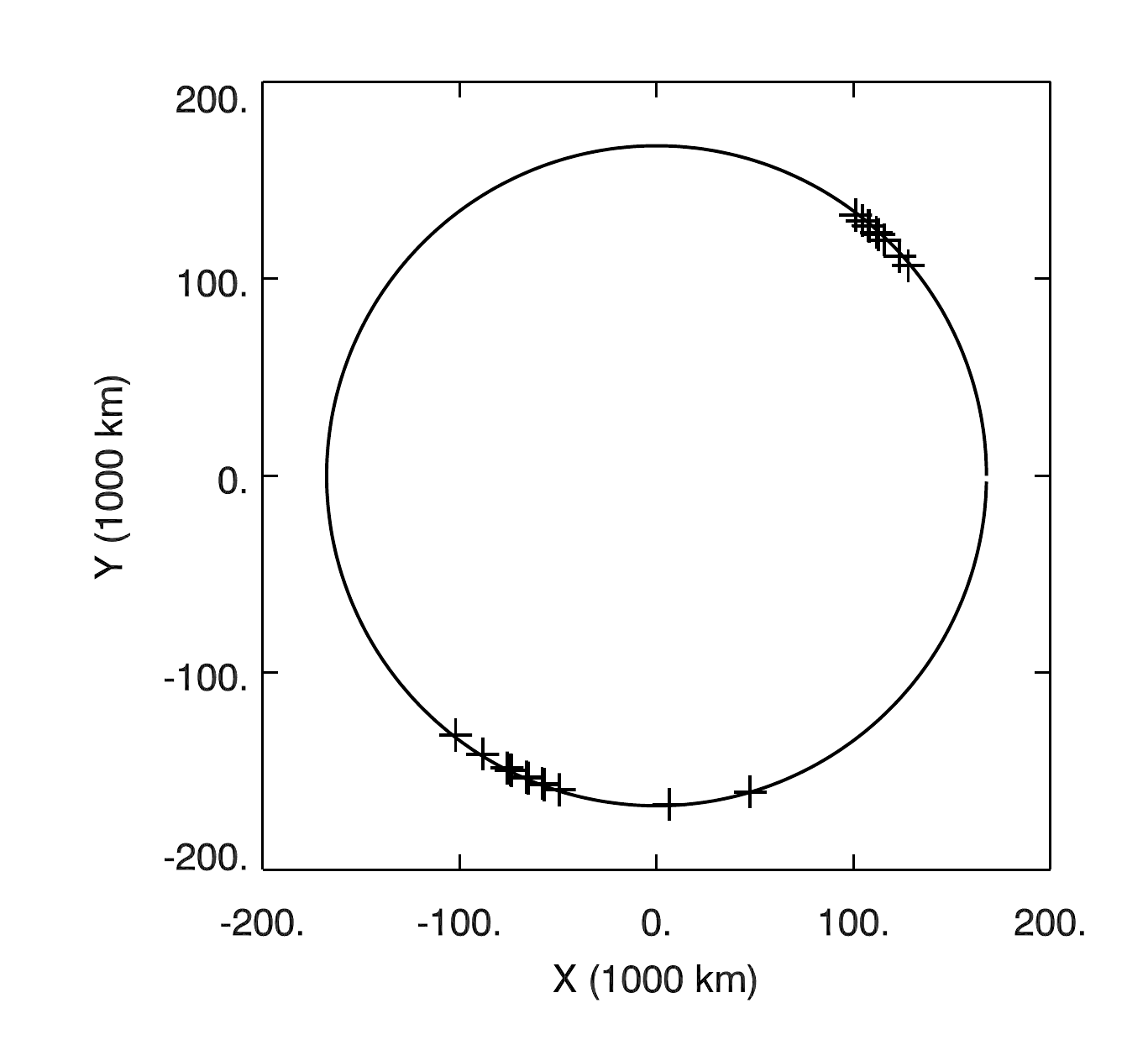}}}
\caption{Observational coverage of Aegaeon projected onto the equatorial plane of Saturn, with superimposed circle of radius 167490 km.}
\label{coverage}
\end{figure}

Numerical integration of both the equations of motion and the variational
equations was performed using the 12th-order Runge-Kutta-Nystr\"om RKN12(10)17M
algorithm of \citet{Brankin89}. For more details on the fitting
procedures, see \citet{Murray05}. The final solution for the state vector at 
epoch in the planetocentric frame, from a fit to the full time-span of 
observations, is given in Table~\ref{solution}. All the observations listed 
in Table~\ref{imgtab} were included in this fit with equal weights. 
Figure~\ref{coverage} shows the orbital coverage of the available observations, 
based on the numerically integrated positions. Note the data fall in two clusters,
which correspond to the two ansae of the orbit when the rings are viewed 
at low phase angles during the observation epoch. 

\begin{figure}[tbp]
\resizebox{6in}{!}{\includegraphics[height=10.0cm]{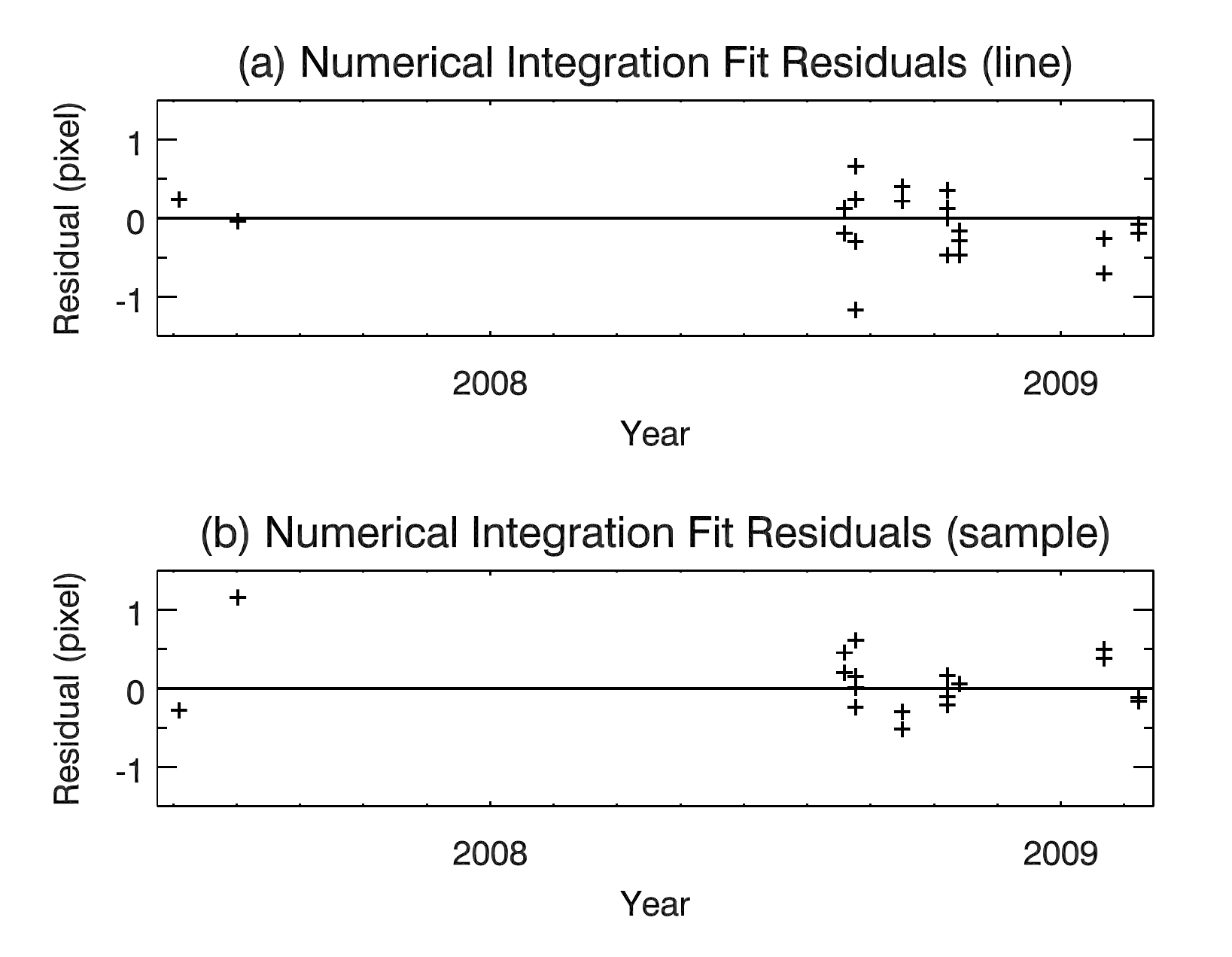}}
\caption{Numerical-integration fit residuals in pixel units : (a) line (b) sample.}
\label{residuals}
\end{figure}

Fit residuals are displayed as a function of time in Fig.~\ref{residuals}.
The overall rms fit residual is 0.468 pixels for the 21 NAC images, which is 
equivalent to 0.578 arcsecond. This is comparable to the residuals for the
NAC observations of Anthe \citep{Cooper08}, which is remarkable given
that in most of the images used here Aegaeon forms a streak several pixels
long. This suggests that our methods for deriving the position of the moon 
are accurate, and that the systematic errors in the modeled orbit are small. 
The final rms uncertainty in the fitted position vector in the frame
of integration is 4.3 km.

Table~\ref{elements} lists the  planetocentric orbital elements derived 
by fitting a uniformly precessing ellipse to the numerically integrated orbit of Aegaeon
over a one-day time-span, using a fine grid of regularly-spaced position vectors. 
These parameters include the semi-major axis $a_{calc}$, eccentricity $e$, 
inclination $i$, longitude of ascending node $\Omega$, longitude of pericenter
$\varpi$, longitude at epoch $\lambda_o$ and mean motion $n$. 
Note that the calculated semi-major axis $a_{calc}$ was obtained by 
first fitting for $n$ and then converting to semi-major axis using the
standard equations involving  Saturn's gravitational harmonics \citep{NP88}. 
The apsidal and nodal rates were calculated using the expressions in
\citet{CM04}, incorporating terms up to $J_6$.
It should be emphasized that the fitted elements in Table~\ref{elements} 
represent only a snapshot of the orbit at the time epoch of fit 
(2008-001T12:00:00 UTC, chosen to be a time when Aegaeon was near the
center of its librations, see below). In reality, the orbital elements
show significant periodic variations due to resonant perturbations from Mimas, 
so a uniformly precessing ellipse will provide a poor approximation of the orbit
over a time span of more than a few days. 

\begin{figure}[htbp]
\resizebox{6in}{!}{\includegraphics{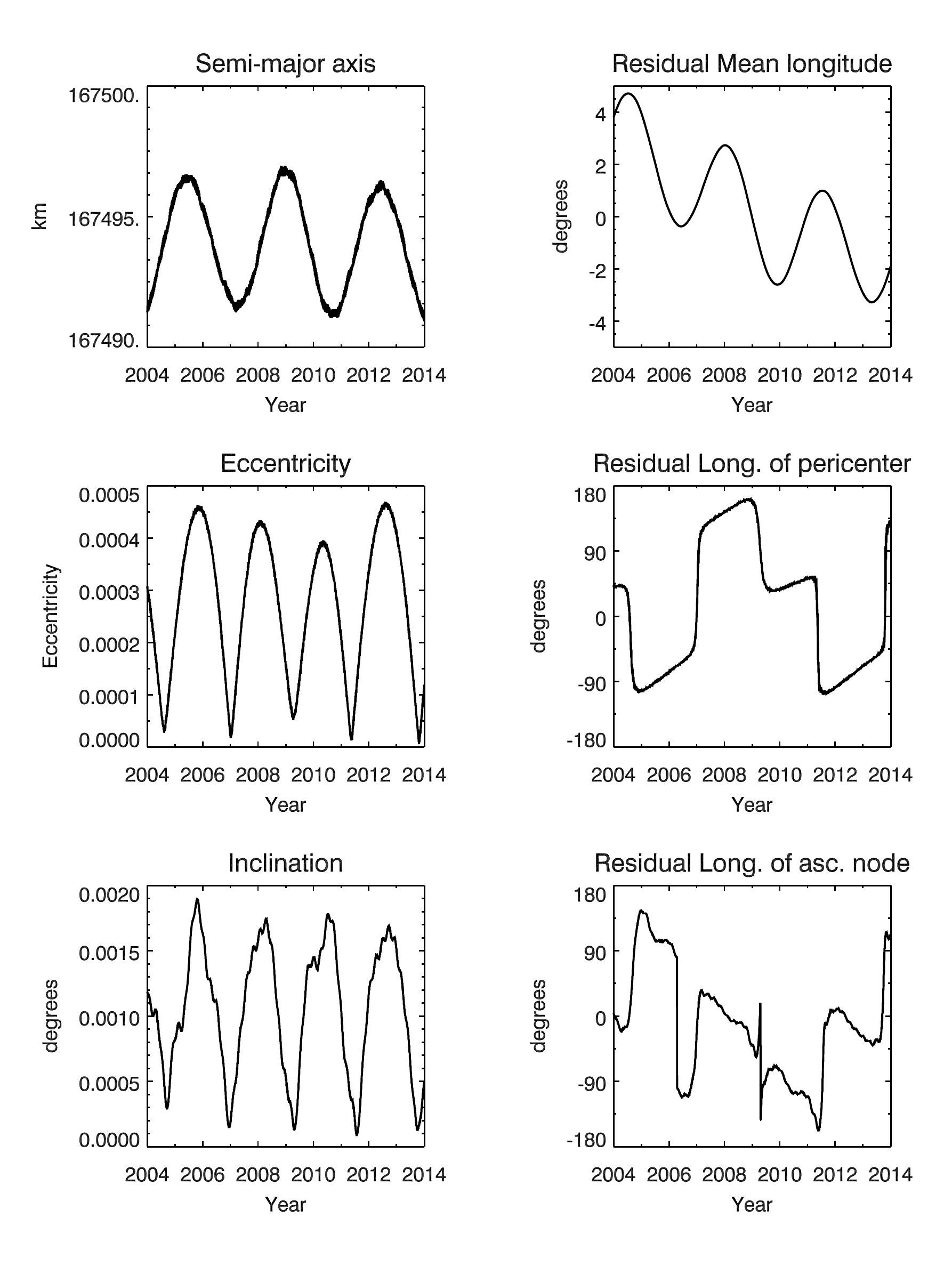}}
\caption{Geometric orbital elements between 2004  and 2014  derived from the numerical integration, including perturbations from the eight major satellites of Saturn plus Prometheus, Pandora, Janus and Epimetheus. Linear background trends have been subtracted from the mean longitude, 
pericenters and nodes prior to plotting (the rates being 445.482$^\circ$/day, 1.146$^\circ$/day
and -1.098$^\circ$/day, respectively).}
\label{elementplot}
\end{figure}

Figure~\ref{elementplot} shows the variations in the geometrical orbital elements
over a period of 10 years.  These plots were generated by integrating the initial
state vector from Table~\ref{solution}, and state vectors were generated at 0.15-day intervals
and converted into geometric orbital elements using standard methods \citep{BL94, Renner04, RS06}.
Unlike conventional osculating elements, these geometric elements are not contaminated 
by short-period terms caused by planetary oblateness.  There  are clear periodic 
variations in all the orbital elements. The semi-major axis varies by 
$\pm$ 4 km around a mean value of 167494  km. The eccentricity ranges
between nearly zero and 0.00047, and the inclination ranges from 0.0001 to 
0.0019 degrees. The mean values of $a$, $e$ and $i$ and the amplitude of their
periodic variations are also given in Table~\ref{elements}. Note that when
the eccentricity and inclination both periodically approach zero, the 
longitudes of node and pericenter change rapidly.

\begin{figure}[tbp]
\centerline{\resizebox{5in}{!}{\includegraphics{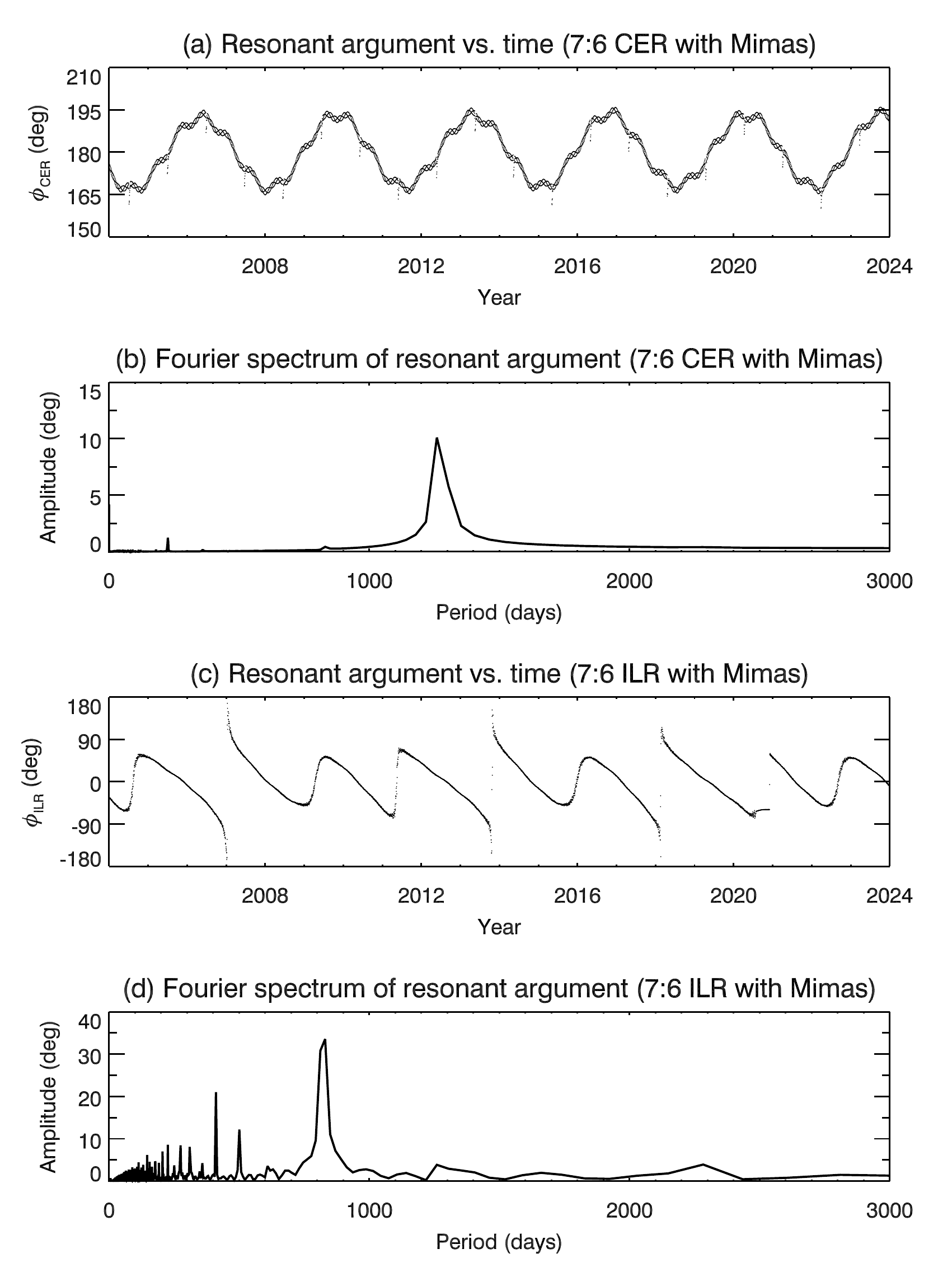}}}
\caption{(a) Resonant argument of the 7:6 CER 
($ \varphi=7\lambda_{Mimas}-6\lambda_{Aegaeon}-\varpi_{Mimas}$) versus time, 
derived from the numerical integration, showing 
that it librates about 180$^\circ$. (b) Fourier spectrum of resonant 
argument, showing a dominant period of approximately 1260 days 
and amplitude 10$^\circ$ (c) Resonant argument of the 7:6 ILR
($\varphi=7\lambda_{Mimas}-6\lambda_{Aegaeon}-\varpi_{Aegaeon}$)
versus time, showing periods of libration around 0$^\circ$
interspersed with brief periods of circulation. (d) Fourier spectrum
of the resonant argument, showing a libration period of approximately 
820 days and amplitude of 35$^\circ$.}
\label{resarg}
\end{figure}

Since the G-ring arc appears to be confined by the 7:6 corotation eccentricity
resonance with Mimas ~\citep{Hedman07}, we expected that Aegaeon would
also be trapped in this resonance. Figures~\ref{resarg}a and b shows the time evolution of the
resonant argument for the 7:6 corotation eccentricity resonance 
\begin{equation}
 \varphi_{CER}=7\lambda_{Mimas}-6\lambda_{Aegaeon}-\varpi_{Mimas}.
 \end{equation}
These data indicate that the argument librates, confirming that Aegaeon 
indeed occupies the 7:6 corotation eccentricity resonance with Mimas.
This analysis also demonstrates that the dominant libration period is 1264 $\pm$ 1 days,
 consistent with the estimated libration periods of particles in the G ring 
 \citep{Hedman07}.

\begin{figure}[tbp]
\resizebox{6in}{!}{\includegraphics{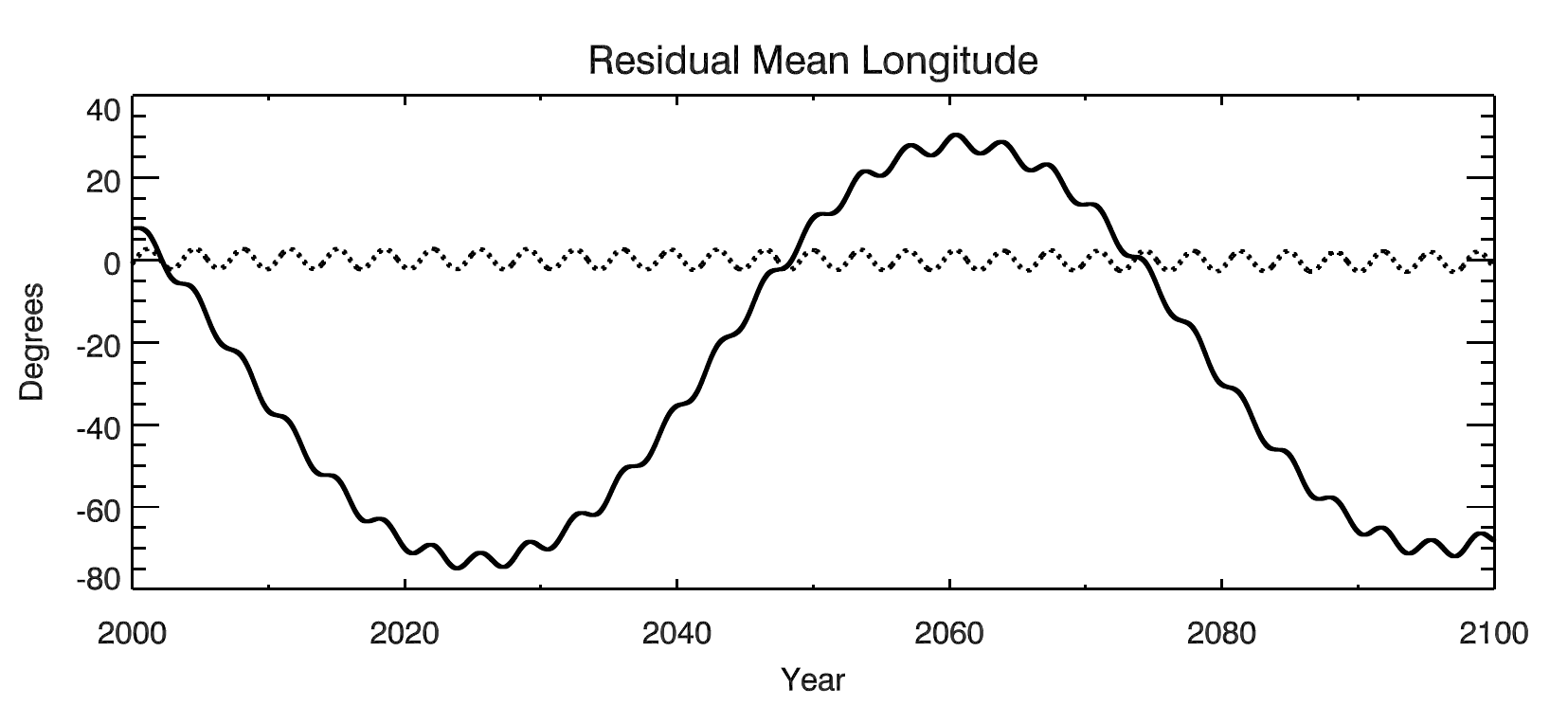}}
\caption{Aegaeon's residual mean longitude (after removing a constant
mean motion of 445.482$^\circ/$day) with and without Tethys
included in the integration. 
Without Tethys, the residual longitude oscillates about zero with an 
amplitude of about 3 degrees, solely due to the effects of the 7:6 CER with 
Mimas (dashed curve). The Mimas:Tethys 4:2 resonance causes the 
additional large amplitude modulation of tens of degrees when Tethys 
is included in the model (solid curve).}
\label{tethys}
\end{figure}

The amplitude of the librations in this resonant argument
is only  $\sim 10^\circ$, so one might expect that 
Aegaeon's longitude would only deviate by a few degrees from
its expected value assuming a constant mean motion. 
In reality, Aegaeon's longitude can drift by tens of degrees
from its expected position assuming a constant mean motion (Fig.~\ref{tethys}).
These long-period drifts have
a characteristic period of 70 years, comparable to the 70.56 year libration
in Mimas' longitude caused by its resonance with Tethys~\citep{VD95}
These variations therefore likely arise because
the longitude of Mimas is itself perturbed
by a resonance with Tethys. Note that 
over the course of the Cassini Mission, the residual longitude
of Aegaeon has drifted backwards 
at  a rate of approximately 0.01$^\circ$/day. This probably 
explains why the best-fit mean-motion of the arc from the Cassini data was
$445.475\pm 0.007^\circ$/day instead of the expected value of 
445.484$^\circ$/day \citep{Hedman07}.

\begin{figure}[tbp]
\centerline{\resizebox{5in}{!}{\includegraphics{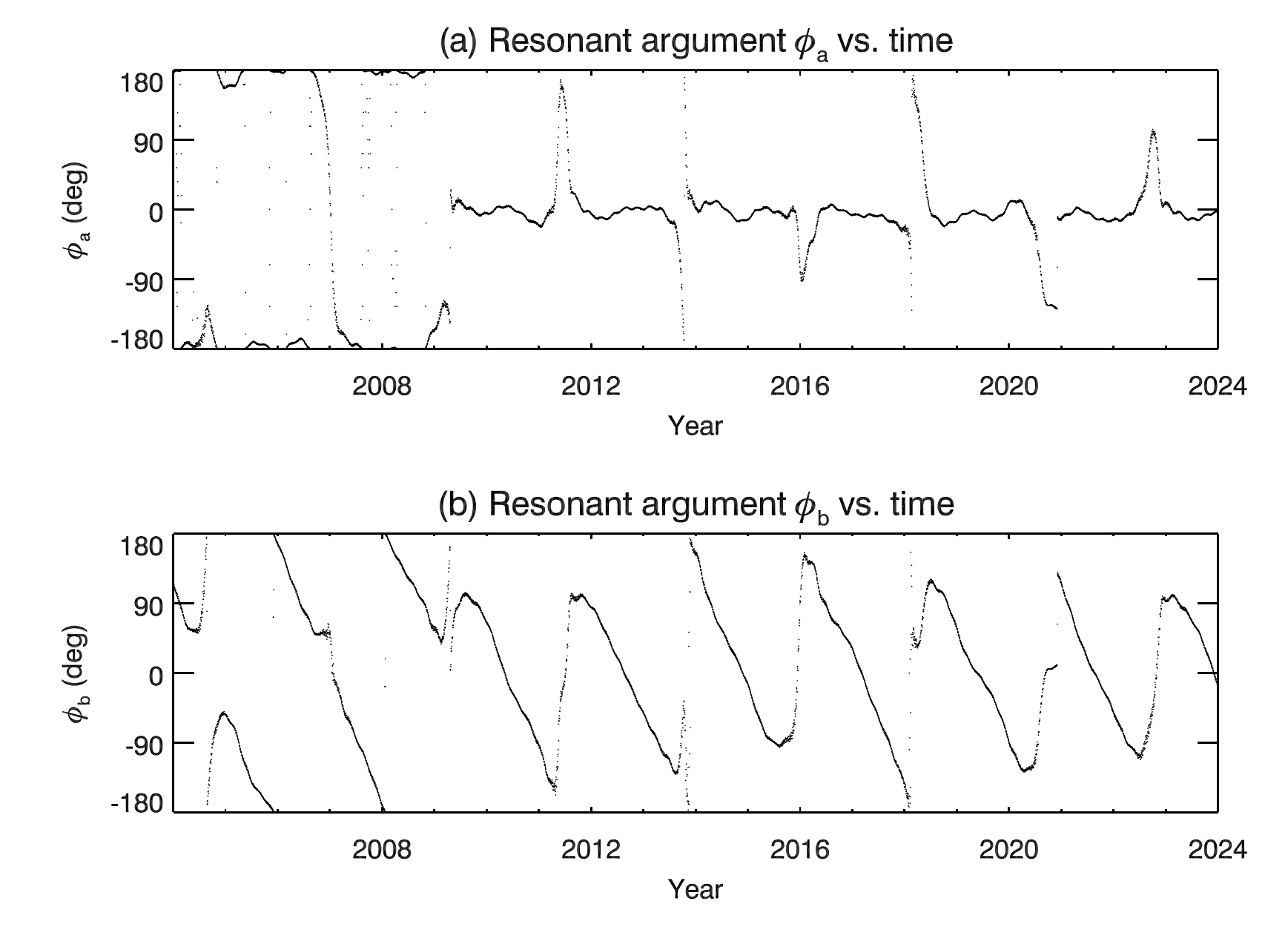}}}
\caption{Resonant arguments $ \varphi_a=7\lambda_{Mimas}-6\lambda_{Aegaeon}
-\varpi_{Aegaeon}-\Omega_{Aegaeon}+\Omega_{Mimas}$ and
$ \varphi_b=7\lambda_{Mimas}-6\lambda_{Aegaeon}
-\varpi_{Aegaeon}+\Omega_{Aegaeon}-\Omega_{Mimas}$versus time, 
derived from the numerical integration.  Note that both these resonant arguments seem
to librate around either $0^\circ$ or $180^\circ$.}
\label{resarg2}
\end{figure}

The corotation eccentricity resonance should primarily affect Aegaeon's orbital
mean motion and computed semi-major axis, and have little 
effect on its eccentricity and inclination. However, there are clearly
large fractional variations in both Aegaeon's eccentricity and inclination.
Furthermore, these variations seem to be coupled, such that
the eccentricity and inclination rise and fall together.  This strongly
suggests that additional resonances are influencing Aegaeon's orbit.
In particular, the correlation between the moon's eccentricity and
inclination suggests a Kozai-like mechanism may be involved. 
However, unlike a classical Kozai Resonance \citep{Kozai62}
where the correlation between the eccentricity and inclination is
negative, in this case the correlation between these two
parameters is positive.

To further explore these aspects of Aegaeon's orbital evolution, 
we looked at the time evolution of the fourteen valid resonant arguments 
to fourth degree in the eccentricities  and the inclinations of the 
form $\varphi=7\lambda_{Mimas}-6\lambda_{Aegaeon}+...$
In addition to the corotation eccentricity resonance, we found 7 other
resonant arguments that exhibited interesting behavior; they are 
listed in Table~\ref{resargtab}. These include the resonant argument 
of the 7:6 Inner Lindblad Resonance $\varphi_{ILR}$, two resonant 
arguments $\varphi_x$ and $\varphi_y$ that can be written as 
linear combinations of $\varphi_{CER}$ and $\varphi_{ILR}$, and 
four resonant arguments involving the nodes of Mimas and Aegaeon
($\varphi_a-\varphi_d$).  Integrations of a few test cases where  the
initial state vector was shifted within the error bars showed the
same fundamental behavior. 
While a detailed investigation of all of these resonant terms is beyond
the scope of this paper, we will briefly discuss the behavior of
a few of these resonant arguments.

Figure~\ref{resarg}c,d shows the time evolution of the resonant
argument of the Inner Lindblad Resonance $\varphi_{ILR}
=7\lambda_{Mimias}-6\lambda_{Aegaeon}-\varpi_{Aegaeon}.$
This resonant argument appears to spend most of its time
librating within $\pm 90^\circ$ of zero with a period of 824 $\pm$ 1
days, interspersed with brief episodes where the resonant argument
circulates around 360$^\circ$. The dominant libration period of this 
argument equals the synodic period of the difference between Aegaeon's
and Mimas' pericenters, which is reasonable since 
$\varphi_{ILR}=\varphi_{CER}+\varpi_{Mimas}-\varpi_{Aegaeon}$.
The alternations between libration and circulation imply that Aegaeon's
orbit lies at the boundary of the ILR, and its free 
eccentricity  is almost equal to the forced eccentricity  
from the Lindblad resonance. As noted previously, the total eccentricity
periodically approaches zero and the pericenter longitude
changes rapidly, as seen in Figure~\ref{elementplot}. During
these episodes, $\varphi_{ILR}$ could either librate through zero
or circulate through $180^\circ$ depending on whether
the forced eccentricity (which will vary with time as Aegaeon's orbit
librates around the corotation eccentricity resonance) 
is slightly larger or smaller than the free eccentricity. 

By way of comparison, it is interesting to note that Saturn's small moon
Methone also appears to occupy both a corotation eccentricity 
resonance and an inner Lindblad resonance
with Mimas \citep{Spitale06, Hedman09}. However, the 14:15 resonances 
occupied by Methone are separated by less than 4 km in semi-major axis, 
while the 7:6 resonances affecting Aegaeon's orbit are separated by 
more than 18 km. The forced eccentricity from the Lindblad
resonance is thus larger for Methone than for Aegaeon, which means 
Aegaeon  needs to have a much smaller free eccentricity to be trapped
in both resonances than Methone does.

\nocite{Morbidelli02}
$\varphi_a-\varphi_d$ appear to be examples of secondary resonances  
i.e. secular resonances existing inside their respective primary mean  
motion resonances. The coupling between Aegaeon's eccentricity and  
inclination mentioned above is also typical of this type of resonant  
motion (for general discussion of secondary resonant behavior, see  
Morbidelli, 2002). Of particular interest are the resonant arguments 
$\varphi_a=\varphi_{CER}+\varpi_{Mimas}+\Omega_{Mimas}-\varpi_{Aegaeon}
-\Omega_{Aegaeon}$ and $\varphi_b=\varphi_{CER}+\varpi_{Mimas}-\Omega_{Mimas}-\varpi_{Aegaeon}+\Omega_{Aegaeon}$. The former is equivalent to a resonant argument
which was found to be librating for Anthe ($\phi_2$ in  Cooper {\it et al.} 2008).
In Aegaeon, this argument typically stays within $\pm 20^\circ$ of either 
$0^\circ$ or $180^\circ$, but can abruptly switch from one state to the
other during periods when the eccentricity and inclination are small (see 
Fig.~\ref{resarg2}a). 
To better understand the significance of this behavior,  note that $\varphi_{CER}$ is already approximately constant,
and that since $\dot{\varpi}_{Mimas} \simeq -\dot{\Omega}_{Mimas}$,
$\varpi_{Mimas}+\Omega_{Mimas}$ is also a constant to good approximation.
Therefore, if $\varphi_a$ remains constant, then $\varpi_{Aegaeon}
-\Omega_{Aegaeon}$ must also be constant, which implies that
$\dot{\varpi}_{Aegaeon} \simeq -\dot{\Omega}_{Aegaeon}$. This is true
for the pericenter precession and nodal regression due to Saturn's oblateness, 
but is not obviously true for the precession and regression due to
perturbations from Mimas. For such perturbations,  the Lagrange equations
indicate that the pericenter
precession rate goes inversely with the eccentricity, while the nodal 
regression rate goes inversely with the inclination. Thus the only
way to have $\dot{\varpi}_{Aegaeon} \simeq -\dot{\Omega}_{Aegaeon}$
is for the eccentricity and inclination vary in step with one another, which
is indeed the case for Aegaeon (see Fig.~\ref{elementplot}). 

$\varphi_b$, like $\varphi_a$, also exhibits periods of circulation and  
libration, although its libration amplitude is far greater than for  
$\varphi_a$ (see Fig.~\ref{resarg2}b), suggesting that Aegaeon is located 
further from the centre of  this particular resonance. This resonant argument
is of particular interest because it can be expressed as
$\varphi_b=\varphi_{CER}+\omega_{Mimas}-\omega_{Aegaeon}$.
$\varphi_b$ is therefore the resonant argument of the CER plus the
difference in the arguments of pericenter of Mimas and Aegaeon, which
suggests that this resonance has some similarities with Kozai
resonances. Although classical Kozai resonances exist  
only at high inclinations, Kozai-type secondary resonances can occur  
inside primary mean motion resonances, even in systems which have  
small eccentricity and inclination \citep{Morbidelli02}. A detailed
investigation of  the secondary resonances represented
by $\varphi_a$, $\varphi_b$, etc. and their implications
for the orbital properties and evolution of Aegaeon is beyond the
scope of this paper, but the number of resonant arguments 
showing interesting behavior indicates that additional 
work on the detailed orbital 
properties of this moon should be quite rewarding..

\section{Comparisons of moon/ring-arc systems}

Aegaeon, like Anthe and Methone, is a small moon embedded
in an arc of debris confined by a first-order corotation eccentricity 
resonance with Mimas.  However Aegaeon also appears to be a 
special case, since it the smallest of these objects while the G-ring arc
is brighter than the arcs surrounding the other moons.  Thus the 
relationship between Aegaeon and the G-ring arc may 
differ from that between the other moons and their
arcs. We therefore compare these systems' dynamical and optical properties.

\begin{figure}[tbp]
\resizebox{6in}{!}{\includegraphics{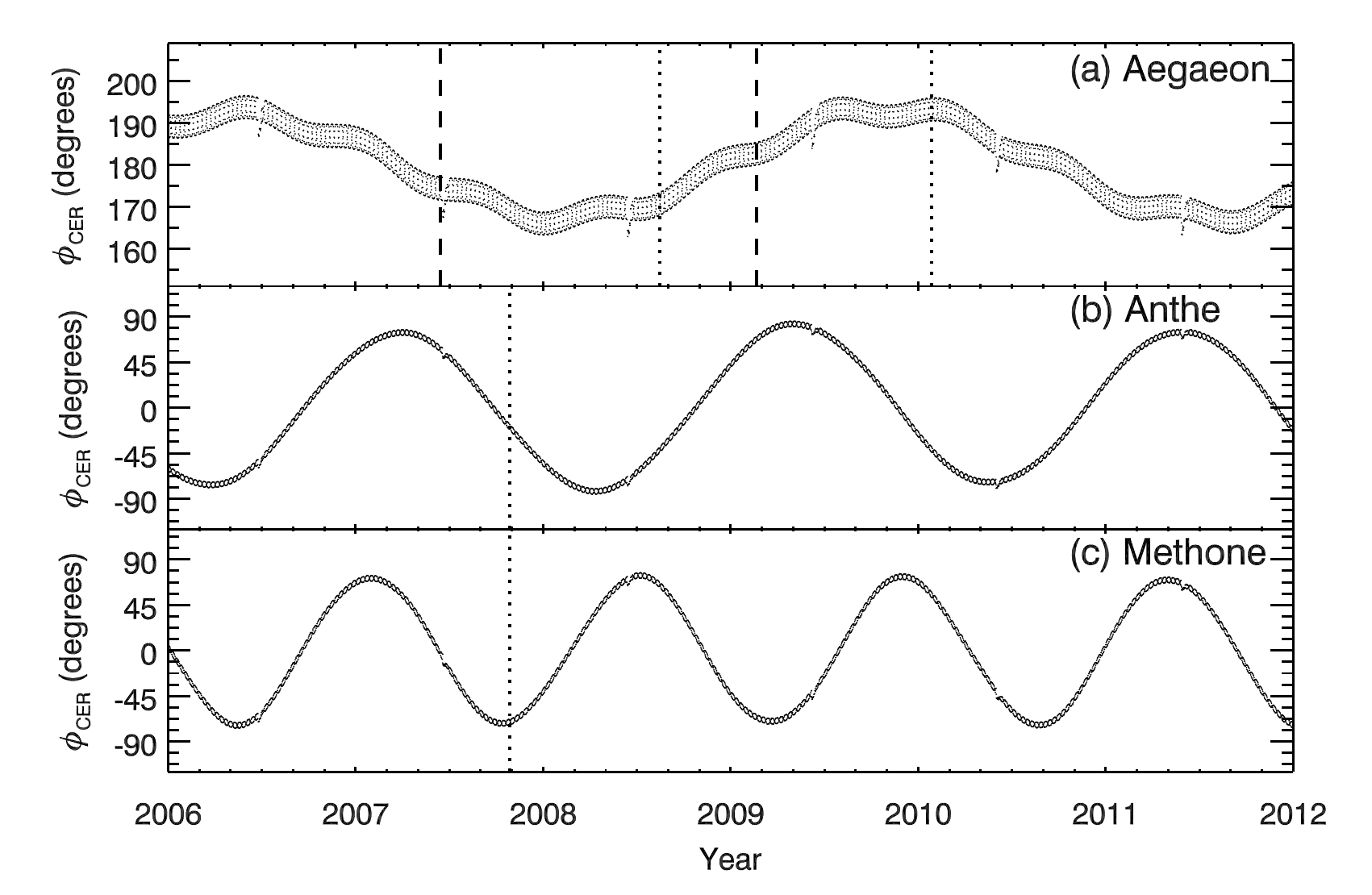}}
\caption{Corotation eccentricity resonant arguments versus time 
for Aegaeon Anthe and Methone. The specific resonant arguments
are (a) Aegaeon $\varphi_{CER}=7\lambda_{Mimas}-6\lambda_{Aegaeon}-\varpi_{Mimas}$ 
(b) Anthe $\varphi_{CER}=11\lambda_{Anthe}-10\lambda_{Mimas}-\varpi_{Mimas}$
and (c) Methone. $\varphi_{CER}=15\lambda_{Methone}-14\lambda_{Mimas}-\varpi_{Mimas}$
In (a) coarse vertical dashed lines 
represent the extent of observational coverage, fine vertical line on 
the right corresponds to 2010-027 and that on the left to 
2008-228. In (b) and (c) vertical dashed lines
 correspond to 2007-302.}
\label{resargs}
\end{figure}

While Aegaeon, Anthe and Methone are all trapped
in corotation eccentricity resonances with Mimas
(7:6, 10:11 and 14:15, respectively), their libration amplitudes 
within those resonances are
quite different. As shown in Figure~\ref{resargs}, the libration 
amplitudes of Anthe and Methone are both between 
70$^\circ$ and $80^\circ$, while the libration amplitude
of Aegaeon is much smaller, only around $10^\circ$.
Aegaeon is therefore more tightly trapped in its
resonance than Anthe and Methone are in theirs.

These differences in the moons' libration amplitudes could explain
some of the differences in the gross morphology
of the various arcs. These morphological differences 
are most visible in longitudinal brightness profiles of the arcs'
radially integrated brightness, which is expressed
in terms of the normal equivalent width:
\begin{equation}
\mathcal{W} = \mu \int{(I/F) dr},
\label{width}
\end{equation}
where $\mu$ is the cosine of the emission angle. Note that
for low optical depth rings, this quantity (with units of length) 
is independent of the viewing geometry and the 
resolution of the images. 

Longitudinal brightness profiles of the Anthe and Methone arcs
were computed in~\citet{Hedman09}, and longitudinal profiles
of the G-ring arc are derived in ~\citet{Hedman07}. However, 
the Anthe and Methone arc profiles are derived
from low-phase-angle ($\sim 23^\circ$) images, while 
the published G-ring arc profiles are derived from high-phase-angle
($>80^\circ$) images, so these published data sets are not 
truly comparable to each other. Fortunately, the same observations
that contain Aegaeon also provide images of the arc at
lower phase angles.  In particular, the
series of images N1597471047-N1597486437 (the sequence in which
Aegaeon was first noticed) captured the entire arc at  
phase angles $\sim 28^\circ$, which is comparable to
the phase angles of the Anthe and
Methone arc observations. A longitudinal brightness profile
of the G-ring arc was derived from these images following
procedures similar to those used in Hedman et al. (2007)
and Hedman et al. (2009). First, the relevant imaging data were 
re-projected onto a grid of radii and longitudes
relative to the predicted location of Aegaeon. To isolate the arc signal
from the rest of the ring, a radial brightness profile of the background
G ring was computed  by averaging the data over
longitudes  between -40$^\circ$ and -50$^\circ$
from Aegaeon, where the arc signal was absent. After subtracting this
background, the normal equivalent width at each longitude
was computed by integrating the brightness over the radial range of 
167,000-168,000 km. 

\begin{figure}[tbp]
\centerline{\resizebox{5in}{!}{\includegraphics{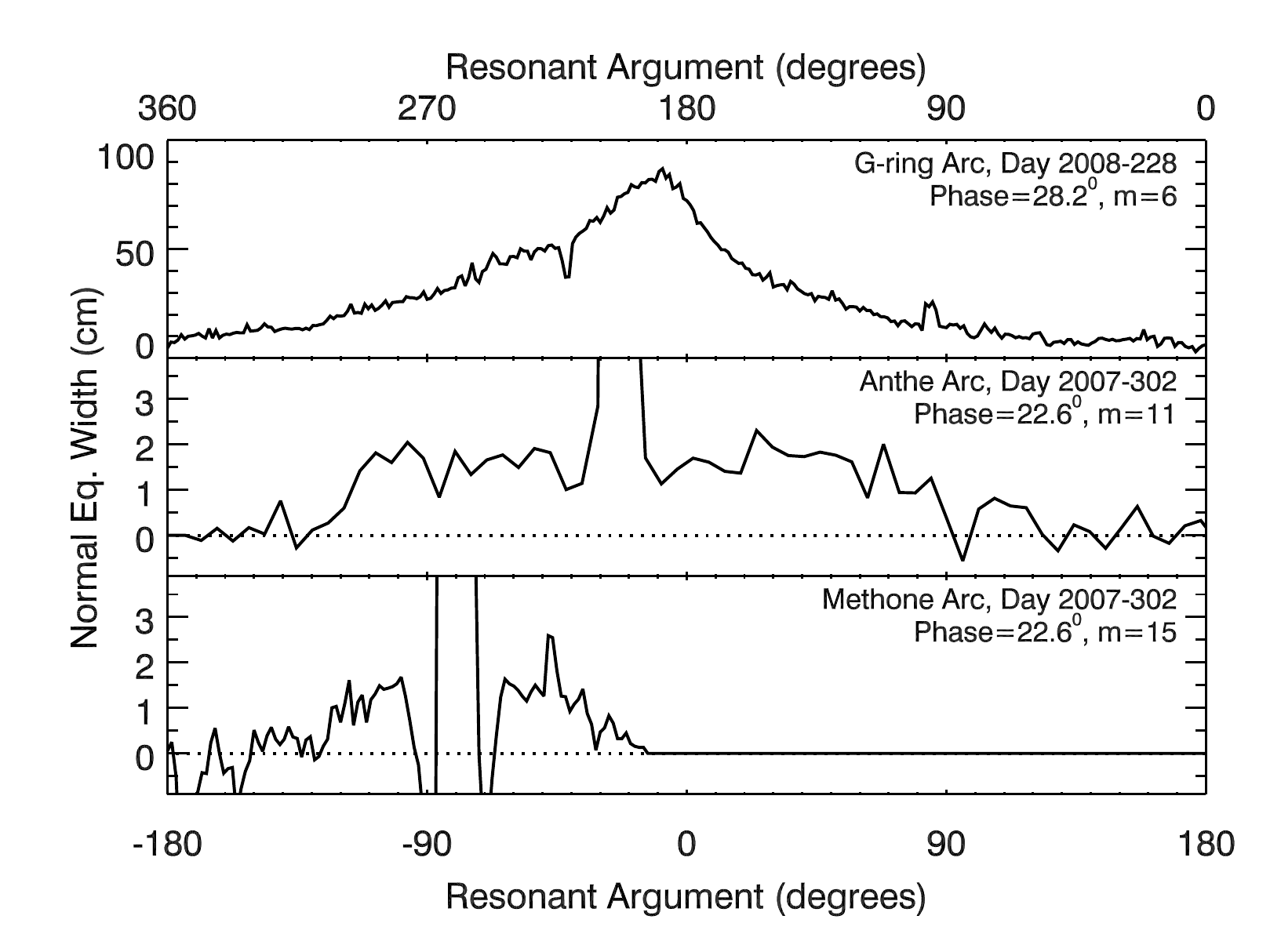}}}
\caption{Longitudinal profiles of the arcs in the G ring (top), Anthe
ring (middle) and Methone ring (bottom). Each profile shows the normal 
equivalent width of the arc versus the appropriate resonant argument of the
appropriate corotation resonance. Note the top axis refers to the G-ring arc
(which is interior to Mimas) while the bottom axis refers to both the Anthe  
and Methone arcs (both of which are exterior to Mimas). In all cases, the
right sides of the plots lead the relevant moons. } 
\label{arcloncomp}
\end{figure}


Figure~\ref{arcloncomp} displays the longitudinal
brightness profiles of the various arcs. Note the $x$-axis
on these plots is the resonant argument $\varphi$ of the appropriate
corotation eccentricity resonances instead of actual 
longitudes, so that the curves can be compared more easily. 
Intriguingly, the G-ring profile has a distinct peak near 
$\varphi=0$, while the Anthe arc is broad with a nearly constant
brightness over a broad range of longitudes. In all likelihood, 
this difference in the morphology of the arcs is directly
related  to the differences in the moons' libration amplitudes 
described above. Aegaeon has a relatively small libration 
amplitude, and so never strays far from $\varphi=0$, while Anthe has a large
libration amplitude and thus moves through a wider range of 
longitudes within the pocket containing the particles. Thus
we might expect that material shed from Anthe would be more evenly distributed
in longitude than the material derived from Aegaeon. Furthermore, 
Anthe should be better able to stir and scatter debris throughout 
the arc as it moves back and forth through the arc. 

The Methone arc presents a more complicated situation, since
the amplitude of Methone's libration is comparable to Anthe's, but
its arc is not as wide. This could possibly be attributed to the fact that 
Methone occupies both the 14:15 corotation resonance and the 
14:15 Lindblad resonance,  and so the dynamics of the particles 
in this region may be more complicated than those in the
Anthe arc.

In addition to the differences in the morphology
of the arcs associated with the various moons, 
the arcs'  overall brightnesses show some interesting
trends. In analogy to the normal equivalent
width given above, one can 
define a normal equivalent area $\mathcal{A}$
as the total integrated brightness of a ring over radius and
longitude $\lambda$:
\begin{equation}
\mathcal{A} = r_o\int \mathcal{W}d\lambda,
\end{equation}
where $r_o$ is the effective mean radius of the ring. Note that this
quantity has units of area and provides a measure of the total
surface area of material in the ring.

Integrating each arc's equivalent width over all longitudes (and 
interpolating the Anthe and Methone arcs over the
region dominated by the signal from the moons), 
we obtain normal equivalent areas of the G-ring, Anthe and
Methone arcs of 50, 1.0 and 0.3 km$^2$, respectively.
The G-ring arc's integrated brightnes is therefore about 2 orders of magnitude 
higher than the arcs associated with Anthe and Methone.
This difference becomes more striking if we compare
these numbers to the effective areas of the moons at comparable 
phase angles. Inserting the values in Table~\ref{moontab} into 
Equation~\ref{phaseeq2}, we find that
the effective areas of Aegaeon, Anthe and Methone at 25$^\circ$ phase
are 0.07, 0.84 and 2.21 km$^2$. For Anthe and Methone, the
normal equivalent areas of the arcs are comparable to the
effective area of the moons, which implies that the debris in these
arcs have comparable surface areas as the moons. Since the particles
in the arcs are likely significantly smaller than the moons, this means 
that the mass in the Anthe and Methone arcs are much less than
the mass in the moons themselves. By contrast, the normal 
equivalent area of the G ring arc is between 10$^3$ and $10^4$
times the effective area of Aegaeon. 

Most of the visible material in the faint
rings likely originates from clouds of debris knocked off the larger 
particles and moons by micrometeoroids, so 
one possible explanation for the distinctive characteristic
of the Aegaeon/G-ring system is that Aegaeon is more efficient
at generating dust than the larger moons Anthe and Methone.
Smaller moons do have lower surface gravity, so a given micrometeoroid 
impact will yield a larger fraction of ejecta that will escape into the ring.
However, smaller moons also have lower cross-sections and
thus have lower impact rates, and theoretical
calculations suggest that the optimal moon size for dust production 
is around 10 km \citep{Burns84, Burns99}. Even though this optimal
size depends on the assumed surface properties of the
source bodies, it is larger than any of the moons considered here, so 
this model predicts that Aegaeon would actually be less
efficient at generating dust than Anthe or Methone.
 
An alternative explanation arises from the realization
that the normal equivalent area of the G ring and the arc
are orders of magnitude higher than the physical area of 
Aegaeon, which is not the case for any of the other ring-moon
systems. Aegaeon therefore does not dominate the cross-section 
of its ring to the same extent as the other moons, so it is quite likely
that Aegaeon shares the arc with a number of other 
objects 1-100 meters across that act as additional sources of
the visible G ring. Such objects would be difficult to see
in the available images because they would
be smeared out into streaks by the long exposure times, which makes them
hard to detect against the background brightness of the G-ring arc. However,  
in-situ measurements provide evidence that additional
source bodies do reside in the G-ring arc. Using in-situ data from
the Voyager spacecraft, \citet{vanAllen83} computed the 
total cross-sectional area of large ($>10$ cm) particles in the G ring
to be 20 km$^2$. This is comparable to the normal equivalent area of the
arc derived above and is much larger than the area of Aegaeon, 
implying that there is indeed a significant population of large objects
in the vicinity of the G ring.  More recently, the MIMI instrument
onboard Cassini detected a $\sim$250-km wide electron 
microsignature associated specifically with the G-ring arc.  
The depth of this microsignature required a total mass of
material equivalent to a roughly 100-meter wide ice-rich moonlet, 
orders of magnitude greater than the mass
in dust-sized grains inferred from images ~\citep{Hedman07}. Furthermore, 
the signature is too wide to be explained by a single moon like Aegaeon, 
which suggests that the arc contains a substantial population of
electron-absorbing source bodies. The G-ring arc therefore appears
 to contain debris with a broad range of sizes,
perhaps the remains of a shattered moon, while the Anthe and Methone
arcs are just the latest small particles knocked off of the relevant moons.

If Aegaeon does share the G-ring arc with a population of
source bodies 1-100 meters across, this could influence 
its dynamics. As Aegaeon librates within the arc, 
it will collide with these smaller objects. Collisions 
within dense arcs of debris confined by corotation resonances
are expected to increase the libration amplitudes of particles
and ultimately allow them to escape the resonance because
collisions dissipate energy and the stable points of corotation
resonances are potential energy maxima \citep{Porco91, Namouni02}. 
However, this situation is slightly different, because we have 
a single large body moving through a sea of
smaller bodies that should have no average net velocity relative to 
the stable point of  the resonance. Hence collisions will act
against any motion of Aegaeon relative to the resonance, and
therefore cause Aegaeon's free inclination, 
free eccentricity, and libration amplitude to decay over time.
 
A crude estimate of the dissipation timescales due to collisions
can be computed by assuming an object of radius $R$
and mass $M$ moves at a velocity $v$ through a background 
medium consisting of a population of small particles.
Say the mass density of the background medium is $\rho_b$,
then the mass encountered by the object in a time $dt$ is 
$\rho_b\pi R^2vdt$. The momentum imparted to this material is 
$C\rho_b\pi R^2v^2dt$, where $C$ is a dimensionless 
constant of order unity.
This must equal the corresponding decrease in the momentum of the object
$Mdv$, so the acceleration of the object due to collisions with the
medium is given by:
\begin{equation}
\frac{dv}{dt}=\frac{C \rho_b\pi R^2v}{M}v.
\end{equation}
Assuming the object has an initial velocity $v_i$ at 
time $t=0$, the velocity will decay 
with time as follows:
\begin{equation}
v(t)=v_i\left(1+\frac{C\pi \rho_b R^2 v_i}{M}t\right)^{-1}.
\end{equation}
Thus the characteristic timescale over which the
velocity falls by a factor of $1/2$ is:
\begin{equation}
t_c=\frac{1}{ C\pi R^2v_i}\frac{M}{\rho_b}.
\end{equation}
Now, since the mass density of the medium (i.e., the arc) is
the most uncertain variable, let us re-express that parameter in terms 
of the arc's mass $m_a$ and its spatial volume $V_a$.
\begin{equation}
t_c=\frac{1}{C}\frac{M}{m_a}\frac{V_a}{\pi R^2v_i}.
\label{timeeq}
\end{equation}
We may now attempt to estimate this characteristic timescale for
moons like Aegaeon and Anthe.
Libration amplitudes of $\sim 10^\circ$  and finite eccentricities of 
$\sim 10^{-3}$ (both reasonable for moons like Aegaeon or Anthe)
lead to typical velocities relative to the resonance's stable point
of order  1 m/s. The G-ring arc has a longitudinal extent 
of $\sim 20^\circ$ or $\sim 6*10^4$ km and a radial width of $\sim$250 km 
\citep{Hedman07}. Assuming its vertical thickness
is comparable to its radial width, the volume of the G-ring arc
is of order $4*10^{18}$ m$^3$. Inserting these numbers
into Equation~\ref{timeeq}, the critical timescale
can be expressed as follows:
\begin{equation}
t_c\simeq\frac{6*10^{5} \mbox{\rm years}}{C}\frac{M}{m_a}
\left(\frac{V_a}{4*10^{18}m^3}\right)\left(\frac{250 m}{R}\right)^2
\left(\frac{1 m/s}{v_i}\right),
\end{equation}
where  all of the terms in parentheses should be of order unity for Aegaeon.

Assuming that Aegaeon has a  mass density of about 
0.5 g/cm$^3$, its mass would be $M \simeq 3*10^{10}$ kg.
Based on the depth of an electron microsignature
observed in the arc's vicinity, \citet{Hedman07}
estimated that the arc's total mass  was between $10^8$
and $10^{10}$ kg (the width of the
microsignature was more consistent with it being associated
with the arc than with the moon). We can therefore estimate
$m_a/M$ to be between $0.003$ and $0.3$, which
would imply damping timescales between 
10$^{6}$ and 10$^{8}$ years. By contrast, Anthe's radius 
is four times larger than Aegaeon's, so its mass is 
$\sim 64$ times larger than Aegaeon's. Furthermore,
assuming the total integrated brightness scales with the
total mass, then the mass of the Anthe arc is at least  $\sim$ 50 
times smaller than that of the G-ring arc.
The characteristic damping time for Anthe should therefore 
be at least $\sim$ 200 times longer than for Aegaeon, 
or $10^{8}$ to $10^{10}$ years. 

Anthe's characteristic damping time is comparable to
the age of the solar system, which implies that collisional
damping has had relatively little effect on Anthe's orbit. Aegaeon's
characteristic damping time is much shorter, so collisional
damping may be significant for this moon. However, the
above values for the damping time will only apply
as long as the moon and the arc have their present
masses. Since hypervelocity impacts with objects on
heliocentric orbits will steadily erode or fragment 
small moons  (cf. Colwell {\it et al.} 2000),
it is likely that Aegaeon was larger in the past than it is today. 
Thus collisional damping can only be effective on 
Aegaeon if its collision damping time is less than
the appropriate erosion or fragmentation time-scale.
 
In lieu of a detailed analysis of Aegaeon's fragmentation history, 
we can roughly estimate how long Aegaeon may have had 
its current size by computing the frequency of catastrophic 
impacts into the moon. The specific energy 
required for catastrophic fragmentation (i.e. the largest remaining
fragment is less than one-half the mass of the original target) 
of an ice-rich object is of order $2*10^{5}$ erg/g (Giblin {\it et al.} 2004,
see also sources cited in Colwell {\it et al.} 2000). Assuming
typical impact velocities of order 40 km/s, this means
catastrophic fragmentation will occur when the ratio of
the impactor's mass to the moon's mass is above about
 $2.5*10^{-8}$. If we again assume that Aegaeon has a 
 mass of about $3*10^{10}$ kg, then
any impactors  with a mass more than 1000 kg would be able
catastrophically disrupt the moon. The present flux of such
objects is quite uncertain, but $10^{-20}$/m$^2$/s is consistent 
with previous estimates and extrapolations \citep{Ip84, CE92}. 
This flux gives a catastrophic impact rate into a 500-m wide Aegaeon 
of order 1 per $10^7$ years. This is comparable to the characteristic
damping time derived above, so these calculations indicate that
Aegaeon could have been  close  to its present size 
over a long enough period of time for collisional damping
to significantly change its orbit. 
Clearly, more detailed analyses are needed to
clarify and quantify the possible interactions between
Aegaeon and the G-ring arc, but these rough calculations
do suggest that collisional damping could provide 
a reasonable explanation for Aegaeon's distinctive 
dynamical properties

\nocite{Colwell00, Giblin04}

\section{Conclusions}

Even though the currently available data on Aegaeon are sparse, 
they are sufficient to demonstrate that it is a interesting object worthy
of further investigation. With a photometrically estimated diameter of
less than a kilometer, Aegaeon is the smallest isolated moon of Saturn yet
observed, and may be comparable in size to the largest particles in Saturn's
main rings, which form the so-called ``giant-propellers" in the A ring \citep{Tiscareno09}. Aegaeon occupies a corotation eccentricity resonance
with Mimas, like Anthe and Methone, and all three of these moons are
associated with resonantly-confined arcs of debris. However, 
Aegaeon also appears to be a special case in terms
of its orbital properties and its relationship with its arc. 
Its eccentricity and inclination are both extremely low, and the large number 
of resonant arguments on the boundary between 
circulation and libration lead to some interesting dynamical behavior. 
At the same time, the mass in the G-ring arc
is probably a significant fraction of (and may even be comparable to) 
Aegaeon's mass, unlike the other arcs associated with small moons, 
opening up the possibility that interactions between the moon and the material 
in the arc could be responsible for some of
Aegaeon's unusual orbital characteristics. Future analysis of
this system could therefore provide insights into the orbital evolution
of satellites coupled to disks of debris.

\section{Acknowledgements}

We thank P.D. Nicholson and J. Veverka for useful conversations, and we also wish to thank two anonymous reviewers for their comments on earlier versions of this manuscript. We acknowledge the support of NASA, the Cassini Project and the Imaging Team. We also wish to thank the VIMS team, who designed several of the observations discussed here. N.C., C.M., K.B. and M.E. acknowledge the financial support
of the UK Science and Technology Facilities Council.

\clearpage

\begin{table}
\caption{Images of Aegaeon}
\label{imgtab}
\resizebox{6in}{!}{\begin{tabular}{|c c c c c c c c c c|}\hline
Image  & Midtime & Range & Phase & B$^b$ & $A_{eff}^c$ & Line$^d$ & Sample$^d$ & RA$^e$  & Dec.$^e$ \\
& (SCET)$^a$ & (km) & (deg.) & (deg.) & (km$^2$) & &  & (deg) & (deg) \\ \hline 
N1560588018  &  2007-166T08:05:49.180 & 1708692 & 42.9 & +0.46 & 0.045 &       521.5 &       396.3 
&191.35064 &+5.1764862\\
N1563866776  &  2007-204T06:51:37.403 & 1432779 & 14.5 &  +0.01 & 0.038 &       500.4 &       523.1 
&60.26121 &+3.2032395\\
N1597476237  &  2008-228T06:45:07.972 & 1188766 & 28.2 & +4.89 & 0.074 &       150.7 &       232.0  
&138.26865 &-4.1721804\\
N1597477967  &  2008-228T07:13:57.959 & 1215278 & 28.2 & +5.14 & 0.072 &       227.9 &       555.0 
&138.32963 &-4.3106590\\
N1598073885  &  2008-235T04:46:01.799 & 1171032 & 12.9 &  -0.75 & 0.072 &       533.4 &       593.5 
&151.92415 &+3.1237490\\
N1598075119  &  2008-235T05:06:35.775 & 1154016 & 13.0 & -0.56 & 0.097 &       745.8 &       619.6 
&151.93067 &+2.7766773\\
N1598104211  &  2008-235T13:11:17.572 & 1179499 & 28.4 & +3.75 & 0.062 &       138.0 &       244.5 
&137.76804 &-3.0640496\\
N1598106121  &  2008-235T13:43:07.559 & 1209028 & 28.4 & +4.02 & 0.074 &       321.2 &       512.2 
&137.80784 &-3.2486713\\
N1600657200  &  2008-265T02:20:48.735 & 1205305 & 15.2 & +4.34 & 0.079 &       659.9 &       377.9 
&153.49124 &-1.8808037\\
N1600659110  &  2008-265T02:52:38.706 & 1177863 & 15.5 & +4.62 & 0.082 &       977.0 &          56.1 
&153.49796 &-2.2741652\\
N1603168767  &  2008-294T04:00:07.953 & 1203664 & 15.0 & +0.09 & 0.082 &       856.5 &       522.8 
&151.72360 &+2.2458309\\
N1603169886  &  2008-294T04:18:46.945 & 1188111 & 14.9 & +0.26 & 0.105 &       576.5 &       525.5  
&151.87725 &+2.0895161\\
N1603171005  &  2008-294T04:37:25.937 & 1172420 & 14.9 & +0.43 & 0.081 &       526.5 &       513.0 
&151.95312 &+1.9205764\\
N1603172124  &  2008-294T04:56:04.929 & 1156757 & 14.9 & +0.60 & 0.077 &       713.8 &       493.2 
&151.94754 &+1.7380537\\
N1603831170  &  2008-301T19:59:56.266 & 1197708 & 30.6 & +4.81 & 0.077 &       254.7 &       414.0 
&137.60681 &-4.0823444\\
N1603831280  &  2008-301T20:01:46.273 & 1199390 & 30.6 & +4.82 & 0.101$^f$&123.4 &       217.6 
&137.61323 &-4.0912850\\
N1603831616  &  2008-301T20:07:22.279 & 1204524 & 30.6 & +4.87 & 0.094$^f$&104.9 &       241.5 
&137.63718 &-4.1180591\\
N1611860574  &  2009-028T18:22:32.246 & 1180159 & 35.0 & -1.49 & 0.072 &       510.1 &       500.8 
&133.96654 &+1.8936283\\
N1611861868  &  2009-028T18:44:06.221 & 1199209 & 34.8 & -1.22 & 0.059 &       273.8 &       403.9 
&134.17857 &+1.6322662\\
N1613784711  &  2009-051T00:51:05.547 & 1186785 & 20.5 & +13.6 & 0.070 &       291.2 &       516.0 
&158.59803 &-10.684674\\
N1613784773  &  2009-051T00:52:28.255 & 1185601 & 20.6 & +13.6 & 0.063 &       289.0 &       475.1 
&158.61153 &-10.715706\\
\hline
\end{tabular}}

\medskip

$^a$ Spacecraft Event Time

\medskip

$^b$ Ring opening angle

\medskip

$^c$ Effective area of the object (see text).

\medskip

$^d$The origin of the image $line$ and $sample$ coordinate system is at the center of the top left pixel, with $line$ increasing downwards and $sample$ to the right, when the image is displayed in its normal orientation. The spacecraft $-$X axis points in the direction of increasing $line$ and $-$Z axes in the increasing $sample$ direction. Estimated measurement uncertainties $\sim$ 0.5 pixel in line and sample. 

\medskip

$^e$RA and DEC refer to right ascension and declination in the International Celestial Reference Frame (ICRF).

\medskip

$^f$Images N1603831280 and N1603831616 taken through RED and IR1 filters, respectively. All other images taken through clear filters. 
\end{table}

\begin{table}
\caption{Images of Pallene}
\label{paltab}
\resizebox{1.9in}{!}{\begin{tabular}{|c r c|}\hline
Image Name & Range & Phase \\
& (km) & (deg.) \\ \hline
N1495207156 & 1140706 &  36.9 \\
N1495207303 & 1141683 &  36.9 \\
N1506004385 & 1619552 &  44.5 \\
N1506004655 & 1616699 &  44.4 \\
N1507534034 & 1396476 &  48.9 \\
N1555052913 & 1703818 &  39.0 \\
N1575630032 & 1564127 &  16.9 \\
N1575675932 & 1878085 &  15.5 \\
N1577009966 & 1748173 &  19.0 \\
N1580356385 & 1362288 &  20.4 \\
\hline
\end{tabular}}
\resizebox{1.9in}{!}{\begin{tabular}{|c r c|}\hline
Image Name & Range & Phase \\ 
& (km) & (deg.) \\ \hline
N1580527536 & 1515543 &  24.8 \\
N1581771720 & 1982140 &  30.8 \\
N1583629498 & 1466714 &  36.8 \\
N1584374373 & 1732141 &  19.9 \\
N1585394936 & 1436059 &  35.9 \\
N1585439051 & 1435253 &  38.5 \\
N1586003505 & 1074505 &  13.7 \\
N1586193031 & 1428772 &  32.7 \\
N1587716367 & 1226979 &  24.4 \\
N1587848623 & 1649791 &  28.8 \\
\hline
\end{tabular}}
\resizebox{1.9in}{!}{\begin{tabular}{|c r c|}\hline
Image Name & Range & Phase \\
& (km) & (deg.) \\ \hline
N1589547370 & 1391832 &  33.2 \\
N1591878927 & 1024706 &  10.2 \\
N1595480632 &   806986 &  25.2 \\
N1595509222 & 1186319 &  30.0 \\
N1597581787 &   999582 &  35.3 \\
N1598065360 &   978305 &  16.7 \\
N1599452540 & 1100556 &  22.2 \\
N1599960489 &   950479 &  18.9 \\ 
N1602671923 & 1049439 &  37.1 \\ 
  & & \\
\hline
\end{tabular}}
\end{table}

\begin{table}
\caption{Images of  Methone}
\label{mettab}
\resizebox{1.9in}{!}{\begin{tabular}{|c r c|}\hline
Image Name & Range & Phase \\
& (km) & (deg.) \\ \hline
N1495209176 & 1254846 &  20.1 \\
N1495209323 & 1252910 &  20.0 \\
N1506063845 &   995105 &  46.7 \\
N1559173514 & 1701162 &  42.3 \\
N1563933254 & 1870692 &  12.2 \\
N1575055798 & 1837767 &  45.6 \\
N1575629432 & 1824208 &  14.1 \\
N1579322353 & 1297453 &  11.2 \\
N1579399258 & 1667738 &  16.5 \\
N1579447529 & 1545970 &  21.5 \\
N1580484276 & 1466609 &  21.8 \\
N1580614807 & 1983810 &  26.2 \\
N1581772425 & 1774910 &  37.2 \\
N1582719892 & 1717981 &  35.3 \\
N1583323256 & 1377304 &  10.5 \\
\hline
\end{tabular}}
\resizebox{1.9in}{!}{\begin{tabular}{|c r c|}\hline
Image Name & Range & Phase \\ 
& (km) & (deg.) \\ \hline
N1583323886 & 1373418 &  10.4 \\
N1583324096 & 1372066 &  10.4 \\ 
N1583344421 & 1179950 &  12.4 \\
N1583757119 & 1646695 &  41.8 \\
N1584374043 & 1651275 &  20.6 \\
N1584714966 & 1559604 &  44.5 \\
N1585394126 & 1577946 &  32.3 \\
N1585438211 & 1363938 &  41.5 \\
N1586002605 & 1389381 &  21.0 \\
N1587747477 & 1539000 &  24.0 \\
N1588451222 & 1361525 &  19.3 \\ 
N1588781885 & 1329843 &  43.1 \\
N1590864949 &   992002 &  38.6 \\
N1591525464 & 1113846 &  35.2 \\
N1591762166 &   945278 &  31.7 \\
\hline
\end{tabular}}
\resizebox{1.9in}{!}{\begin{tabular}{|c r c|}\hline
Image Name & Range & Phase \\ 
& (km) & (deg.) \\ \hline
N1591878207 & 1211541 &    9.3 \\ 
N1595481232 &   832264 &  26.4 \\
N1595510182 & 1239193 &  26.8 \\
N1596877322 & 1037741 &  25.9 \\
N1597581487 & 1026605 &  35.1 \\
N1598065750 & 1286329 &  13.5 \\ 
N1599961704 & 1318799 &  26.9 \\
N1600651648 & 1289847 &  29.5 \\
N1600751290 & 1357620 &  28.2 \\
N1601291954 & 1344186 &  16.2 \\ 
N1601855958 &   994265 &   31.2 \\
N1602578562 & 1370014 &  27.4 \\
N1603214386 & 1309285 &  16.7 \\
N1604534936 & 1226594 &  21.5 \\
N1604570261 & 1281208 &  33.8 \\
\hline
\end{tabular}}
\end{table}

\begin{table}
\caption{Images of  Anthe}
\label{anttab}
\resizebox{1.9in}{!}{\begin{tabular}{|c r c|}\hline
Image Name & Range & Phase \\
& (km) & (deg.) \\ \hline
N1572352978 & 2304579 &  22.6 \\
N1572353038 & 2303910 &  22.6 \\
N1572353098 & 2303241 &  22.6 \\
N1572353158 & 2302571 &  22.6 \\
N1572353218 & 2301901 &  22.6 \\
N1572353442 & 2299328 &  22.7 \\ 
N1575629162 & 1802824 &  14.3 \\
N1579321873 & 1440693 &  11.3 \\
N1579364158 & 1258476 &  17.9 \\
N1580356175 & 1214702 &  20.0 \\
N1581514393 & 1472750 &  21.2 \\
\hline
\end{tabular}}
\resizebox{1.9in}{!}{\begin{tabular}{|c r c|}\hline
Image Name & Range & Phase \\ 
& (km) & (deg.) \\ \hline
N1582636683 & 1783724 &  27.7 \\
N1582768099 & 1419578 &  41.9 \\
N1583627560 & 1746544 &  33.9 \\
N1585394528 & 1619129 &  31.4 \\
N1586002250 & 1235543 &  23.5 \\
N1586004500 & 1271592 &  23.4 \\
N1591878477 & 1065302 &    9.7 \\
N1596338308 & 1242828 &  28.9 \\
N1596721036 &   861867 &  21.8 \\
N1598065060 &   990740 &  17.2 \\
N1599961164 & 1029567 &  30.0 \\
\hline
\end{tabular}}
\resizebox{1.9in}{!}{\begin{tabular}{|c r c|}\hline
Image Name & Range & Phase \\ 
& (km) & (deg.) \\ \hline
N1600583458 &   965664 &  28.8 \\
N1600749010 & 1034862 &  27.2 \\
N1601479534 &   854848 &  48.2 \\
N1601778096 & 1097628 &  34.8 \\
N1601856348 & 1202995 &  32.6 \\
N1602516863 & 1273685 &  15.0 \\
N1602577242 & 1331594 &  29.3 \\
N1603720951 &   935528 &  36.3 \\
N1603880961 & 1059788 &  22.0 \\
N1604739758 & 1061249 &  44.8 \\
& & \\
\hline
\end{tabular}}
\end{table}

\begin{table}
\caption{Summary of photometric properties of the small moons}
\label{moontab}
\begin{tabular}{|c|ccc|}\hline
Moon & $\beta$ & $p_{eff}<A_{phys}>$ & $<r_{phys}>^a$ \\ \hline
Pallene    & 0.017 mag/degree & 7.38 km$^2$   & 2.2 km \\
Methone  & 0.023 mag/degree & 3.76 km$^2$   & 1.6 km \\
Anthe       & 0.032 mag/degree & 1.76 km$^2$   & 1.1 km \\
Aegaeon$^b$ & 0.007 mag/degree & 0.084 km$^2$ & 0.24 km \\ \hline
\end{tabular}

\medskip

$^a$ Assuming all four moons have $p_{eff}=0.49$
(required to match mean radius of Pallene).

\medskip

$^b$ Fit to only the $|B|>1^\circ$ data.
\end{table}

\begin{table}
\caption{Saturn constants used in orbit fitting and numerical modeling}
\label{satparams}
\begin{tabular}{|lll|}\hline
\hline
Constant&Value$^a$&units \\ \hline
 Pole (RA,Dec)&  (40.5837626692582, 83.5368877051669) &deg\\
 $GM$  &     37931207.1585 &km$^3$ s$^{-2}$ \\
 Radius, $R_s$          &60330  &       km\\
 $J_{2}$         & 0.016290543820& \\
 $J_{4}$         &$-$0.000936700366& \\
 $J_{6}$         &0.000086623065& \\ \hline
\end{tabular}

\medskip
$^a$Pole position from SPICE kernel cpck19Sep2007.tpc, precessed to the fit epoch. Reference radius from \citet{Kliore80}. Zonal harmonics and $GM$ from cpck19Sep2007.tpc 
\end{table}

\begin{table}
\caption{SPICE kernels used in orbit	 fitting and numerical modeling}
\label{kernels}
\centerline{\resizebox{2in}{!}{\begin{tabular}{|l|}\hline
Kernel name$^a$ \\ \hline
pck00007.tpc\\
naif0009.tls\\
cas00130.tsc\\
cpck19Sep2007.tpc\\
cpck\_rock\_01Oct2007\_merged.tpc\\
de414.bsp\\
jup263.bsp\\
sat286.bsp\\
080806AP\_SCPSE\_08138\_10182.bsp\\
081211AP\_SCPSE\_08346\_08364.bsp\\
090120AP\_SCPSE\_09020\_09043.bsp\\
090202BP\_SCPSE\_09033\_09044.bsp\\
090209AP\_SCPSE\_09037\_09090.bsp\\
090305AP\_SCPSE\_09064\_09090.bsp\\
081125AP\_RE\_90165\_18018.bsp\\
070416BP\_IRRE\_00256\_14363.bsp\\
070727R\_SCPSE\_07155\_07170.bsp\\
070822R\_SCPSE\_07170\_07191.bsp\\
071017R\_SCPSE\_07191\_07221.bsp\\
071127R\_SCPSE\_07221\_07262.bsp\\
080117R\_SCPSE\_07262\_07309.bsp\\
080123R\_SCPSE\_07309\_07329.bsp\\
080225R\_SCPSE\_07329\_07345.bsp\\
080307R\_SCPSE\_07345\_07365.bsp\\
080327R\_SCPSE\_07365\_08045.bsp\\
080428R\_SCPSE\_08045\_08067.bsp\\
080515R\_SCPSE\_08067\_08078.bsp\\
080605R\_SCPSE\_08078\_08126.bsp\\
080618R\_SCPSE\_08126\_08141.bsp\\
080819R\_SCPSE\_08141\_08206.bsp\\
080916R\_SCPSE\_08206\_08220.bsp\\
081031R\_SCPSE\_08220\_08272.bsp\\
081126R\_SCPSE\_08272\_08294.bsp\\
081217R\_SCPSE\_08294\_08319.bsp\\
090120R\_SCPSE\_08319\_08334.bsp\\
090202R\_SCPSE\_08334\_08350.bsp\\
090225R\_SCPSE\_08350\_09028.bsp\\
090423R\_SCPSE\_09028\_09075.bsp\\
\hline
\end{tabular}}}

\medskip

$^a$Kernels are available by anonymous ftp from
{\tt  ftp://naif.jpl.nasa.gov/pub/naif/CASSINI/kernels}
\end{table}

\begin{table}
\caption{$GM$ values for other perturbing bodies used in orbit fitting and numerical modeling}
\label{pertparams}
\centerline{\begin{tabular}{|lc|} \hline
Body& $GM^a$ (km$^3$ s$^{-2}$) \\ \hline
Sun &  132712440044.2083\\
Jovian system & 126712764.8582231 \\
Prometheus & 0.01058 \\
Pandora & 0.00933 \\
Janus & 0.12671 \\
Epimetheus & 0.03516 \\
Mimas & 2.50400409891677 \\
Enceladus & 7.20853820010930 \\
Tethys & 41.2103149758596 \\
Dione & 73.1128918960295 \\
Rhea & 153.941891174579 \\
Titan& 8978.13867043253 \\
Hyperion & 0.370566623898283 \\
Iapetus & 120.504895547942 \\ \hline
\end{tabular}}
$^a$Values from SPICE kernels cpck19Sep2007.tpc and cpck\_rock\_10Oct2007\_merged.tpc.
\end{table}

\begin{table}[hp]
\caption{Solution for the planetocentric state vector of Aegaeon, from a fit to $Cassini$ ISS data in 
the ICRF frame. Epoch is 2008-228T06:45:07.972 UTC (2008-228T06:46:13.154 or JD 2454693.78209670 TDB)}
\label{solution}
\begin{center}
\begin{tabular}{|c|l|c|}
\tableline
Aegaeon&&Units \\
\tableline
$x$&±0.123456139640300E+06 $\pm$3.2855191576&km\\
$y$&±0.111538222264526E+06 $\pm$2.4140374161&km\\
$z$& $-$0.188394292598307E+05 $\pm$1.2178065755&km\\
$\dot{x}$& $-$0.101036088689253E+02 $\pm$0.0003052052&km\,s$^{-1}$\\
$\dot{y}$&±0.111912435677253E+02 $\pm$0.0001445381&km\,s$^{-1}$\\
$\dot{z}$&±0.447482819944934E-01$\pm$0.0001146560&km\,s$^{-1}$\\
\tableline
rms& 0.468 & pixel\\
rms& 0.578 & arcsec\\
\tableline
\end{tabular}
\end{center}
\end{table}

\begin{table}[hp]
\caption{Planetocentric Orbital Elements}
\label{elements}
\begin{tabular}{|l|l|l|}
\tableline
Parameter$^a$ &Fitted Value&Units \\
\tableline
 Fit Epoch  & 2008-001T12:00:00.000 UTC&UTC\\
  & 2008-001T12:01:05.183 TDB & TDB \\
  & JD 2454467.00075444 TDB & TDB \\
 Semi-major axis, $a_{\rm calc}$&167493.665 $\pm$ 0.004& km\\
 Eccentricity, $e$ &0.00042277 $\pm$ 0.00000004 &\\
 Inclination, $i$ &0.0007$\pm$ 0.6& deg\\
 Longitude of ascending node, $\Omega$ &30 $\pm$ 298& deg\\
 Longitude of pericenter,$\varpi$ &352.694 $\pm$ 0.005& deg\\
 Mean longitude, $\lambda$&45.606789 $\pm$ 0.000004& deg\\
 Mean motion, $n$  &445.48328 $\pm$ 0.00002& deg/day\\
 Pericenter rate, $\dot{\varpi}_{\rm calc}$ &1.43691010& deg/day\\
 Nodal rate, $\dot{\Omega}_{\rm calc}$ &$-$1.43229434& deg/day\\
\hline
\hline
Parameter & Mean Value $\pm$ Libration & Units \\
\tableline
 Semi-major axis, $a_{\rm mean}$ &167494 $\pm$ 4& km\\
 Eccentricity, $e_{\rm mean}$ &0.00024 $\pm$ 0.00023&\\
 Inclination, $i_{\rm mean}$  & 0.0010 $\pm$ 0.0009& deg\\
 Resonant argument (CER) &\begin{math}7\lambda_{Mimas}-6\lambda_{Aegaeon}-\varpi_{Mimas} \end{math} & \\
 Resonant argument's libration period (CER) & 1264 $\pm$ 1 & days\\
 Resonant argument (ILR) &\begin{math}7\lambda_{Mimas}-6\lambda_{Aegaeon}-\varpi_{Aegaeon} \end{math} & \\
 Resonant argument's libration period(ILR) & 824 $\pm$ 1 & days\\

\tableline
\end{tabular}

\medskip

$^a$All longitudes measured directly from ascending node of Saturn's equator on 
the ICRF equator. Inclination measured relative to Saturn's equatorial plane. Quoted uncertainties in the upper half of the table are the formal 1$\sigma$ values from the fit. Note the values in the upper half of the table are values at a particular point in time and are provided as a guide. They are not suitable as starting conditions in integrations for the equation of motion. Mean values and their librations in the lower half of the table were obtained from a numerical integration of the period 2004-001 to 2014-001, taking into account the resonant behavior. 

\end{table}

\begin{table}
\caption{Resonant Arguments of Interest}
\label{resargtab}
\resizebox{6in}{!}{\begin{tabular}{|l|l|l|}\hline
Name & Argument & Argument in terms of $\varphi_{CER}$ and $\varphi_{ILR}$ \\ \hline
$\varphi_{CER}$ & $7\lambda_{Mimas}-6\lambda_{Aegaeon}-\varpi_{Mimas}$  & $\varphi_{CER}$ \\ \hline
$\varphi_{ILR}$ & $7\lambda_{Mimas}-6\lambda_{Aegaeon}-\varpi_{Aegaeon}$ &  
$\varphi_{CER}+\varpi_{Mimas}-\varpi_{Aegaeon}$	\\ \hline
$\varphi_x$ & $7\lambda_{Mimas}-6\lambda_{Aegaeon}+\varpi_{Mimas}-2\varpi_{Aegaeon}$ & 
$\varphi_{CER}+2\varpi_{Mimas}-2\varpi_{Aegaeon}=2\varphi_{ILR}-\varphi_{CER}$ \\
$\varphi_y$ & $7\lambda_{Mimas}-6\lambda_{Aegaeon}-2\varpi_{Mimas}+\varpi_{Aegaeon}$ &
$\varphi_{CER}-2\varpi_{Mimas}+2\varpi_{Aegaeon}=2\varphi_{CER}-\varphi_{ILR}$ \\ \hline
$\varphi_a$ & $7\lambda_{Mimas}-6\lambda_{Aegaeon}-\varpi_{Aegaeon}
	+\Omega_{Mimas}-\Omega_{Aegaeon}$ & 
		$\varphi_{CER}+\varpi_{Mimas}+\Omega_{Mimas}
			-\varpi_{Aegaeon}-\Omega_{Aegaeon}$ \\
& &   =$\varphi_{ILR}+\Omega_{Mimas}-\Omega_{Aegaeon}$ \\
$\varphi_b$ & $7\lambda_{Mimas}-6\lambda_{Aegaeon}-\varpi_{Aegaeon}
	-\Omega_{Mimas}+\Omega_{Aegaeon}$ & 
$\varphi_{CER}+\varpi_{Mimas}-\Omega_{Mimas}
			-\varpi_{Aegaeon}+\Omega_{Aegaeon}$ \\
& & =$\varphi_{CER}+\omega_{Mimas}-\omega_{Aegaeon}$ \\
& & =$\varphi_{ILR}-\Omega_{Mimas}+\Omega_{Aegaeon}$ \\
$\varphi_c$ & $7\lambda_{Mimas}-6\lambda_{Aegaeon}-\varpi_{Mimas}
	+\Omega_{Mimas}-\Omega_{Aegaeon}$ &
$\varphi_{CER}+\Omega_{Mimas}
			-\Omega_{Aegaeon}$ \\ 
$\varphi_d$ & $7\lambda_{Mimas}-6\lambda_{Aegaeon}-\varpi_{Mimas}
	-\Omega_{Mimas}+\Omega_{Aegaeon}$ &
		$\varphi_{CER}-\Omega_{Mimas}
			+\Omega_{Aegaeon}$ \\ \hline
\end{tabular}}
\end{table}

\clearpage


\end{document}